\documentstyle[onecolumn,epsfig]{mn2e}
\newcommand{\mincir}{\raise
-2.truept\hbox{\rlap{\hbox{$\sim$}}\raise5.truept
\hbox{$<$}\ }}
\newcommand{\magcir}{\raise
-2.truept\hbox{\rlap{\hbox{$\sim$}}\raise5.truept
\hbox{$>$}\ }}
\newcommand{\minmag}{\raise-2.truept\hbox{\rlap{\hbox{$<$}}\raise
6.truept\hbox{$>$}\ }}
\newlength{\royalength}
\newcommand{\be}{\begin{equation}}
\newcommand{\ee}{\end{equation}}
\newcommand{\ba}{\begin{eqnarray}}
\newcommand{\ea}{\end{eqnarray}}
\newcommand{\brr}{\begin{array}}
\newcommand{\nn}{\nonumber}
\newcommand{\err}{\end{array}}
\newcommand{\bc}{\begin{center}}
\newcommand{\ec}{\end{center}}

\newcommand{\ie}{i.e.~}

\begin{document}

\title[Gravitational cooling and density profile near caustics in 
collisionless haloes]{Gravitational cooling and density profile
near caustics in collisionless dark matter haloes}
\author[Roya Mohayaee \& Sergei Shandarin]
{Roya Mohayaee$^*$ and Sergei F. Shandarin$^\dagger$\\
$^*$Institut d'Astrophysique de Paris, 98bis boulevard Arago, 75014 Paris, France \\
$^\dagger$Department of Physics and Astronomy, University 
of Kansas, KS 66045, U.S.A.\\
emails: roya@iap.fr\, ;\,  sergei@ku.edu}
%\date{Accepted 2000 ???? ???; Received 2000???? ???;
%in original form 2000???? ??}
%\date{\bf Preliminary version}

%\begin{document}

\maketitle

\begin{abstract}
Cold dark matter haloes are populated by high-density
structures with sharply-peaked profiles known as caustics which 
have not yet been resolved by 3-dimensional numerical simulations.
Here, we derive semi-analytic expressions for the density profiles
near caustics in haloes which form by self-similar accretions 
of dark matter with infinitesimal velocity dispersion.
A simple rescaling shows that similarly to the case of absolutely cold medium
these profiles  are universal: they are valid for all caustics
and irrespective of physical parameters of the halo.
We derive the maximum density 
of the caustics and show that it depends 
on the velocity dispersion and the caustic 
location. We show that both the absolute and relative thickness
of the caustic  monotonically decrease toward the center
of the halo while the maximum density grows. 
This indicates that the radial component of the thermal velocities 
decreases in the inner streams, \ie {\it the collisionless  medium cools down 
in the radial direction descending to the center of the halo}. 
Finally, we demonstrate that there can be a significant
contribution to the emission measure from dark matter particle 
annihilation in the caustics.

\end{abstract}


%!!!!!!!!!!!!!!!!!!!!!!!!!!!!!
\section{Introduction}
\label{sec:introduction}
%!!!!!!!!!!!!!!!!!!!!!!!!!!!!!

Dark matter particles, if collisionless and cold, 
would focus under gravitational instability into {\it caustics}
which are formally 2D manifolds of infinite density. 
In three-dimensional space, caustics are determined by 
tangent hyperplanes {\bf r} = {\it const}
to the phase surface in the six-dimentional phase space ({\bf r}, {\bf v})
and bound the regions of {\it
multi-stream} flow where velocity has multiple values. 
Once formed, caustics of a given phase volume 
neither disappear nor overlap; a requirement
of the Liouville theorem. However they can
interact and merge with caustics of 
a different phase volume and consequently the nature of their
singularity could change and they could undergo generic {\it metamorphoses}. 
In the case of potential flows, {\it e.g.} of light rays or 
cold dark matter on large scales, the singularities of the caustics and their
metamorphosis have been classified up to three spatial dimensions 
(Arnol'd 1986, 1990).
This classification remains intact in the presence of external or
internal forces (e.g. in a self-gravitating system) for as long as the force is
potential and smooth and the dark matter can be approximated as a collisionless fluid. The first and most common caustic has a
density with an inverse square-root singularity (as occurs in the
Zel'dovich approximation). This singularity has been rigorously proven to be robust in
the case of a one-dimensional Vlassov-Poisson system (Roytvarf 1994) and is the only singularity that is of relevance to the present work.  The other generic singularities
can lie on one-dimensional manifolds (lines) or be isolated points.  
Generally, one-dimensional singularities are stronger than two-dimensional 
singularities and zero-dimensional are stronger
than one-dimensional singularities. None of these singularities 
requires any particular symmetry for its formation
(Arnol'd, Shandarin, Zel'dovich 1982, Arnol'd, Gusein-Zade, Varchenko 1985,
Shandarin, Zel'dovich 1989).

In cosmology, the study of formation and evolution of dark matter caustics
has been historically and scientifically twofold: caustic related to the large-scale
structure and caustic on the galaxy or smaller scales. On large scales, 
pioneering works by Zel'dovich and his collaborators 
(Zel'dovich 1970, Arnol'd, Shandarin and Zel'dovich 1982, Shandarin and Zel'dovich 1989) showed that 
in a suitably-defined time and space coordinates, elements of a collisionless 
and self-gravitating fluid
move on inertial trajectories: {\it i.e.} with their initial velocities. Thus,
as in ray optics, when the path of the free-moving particles cross, density
diverges, velocity becomes multivalued and caustics form. 
Caustics exist only as idealizations in models assuming that the medium is
collisionless, continuous and cold \ie the thermal velocity dispersion equals zero.
It was shown that in the case of finite thermal velocity dispersion the density
in the caustic regions becomes finite (Zel'dovich, Shandarin 1982).
Although the real thermal velocity dispersion is never zero in many cases it is 
extremely small and
the cold medium  represent an excellent first approximation to reality.
Thus, caustics of various types represent  a very useful idealization for the study
of complex density fields.  It needles to say that both discreteness and 
collisionallity eliminate caustics.
Caustics can bound regions with 
the morphology of filaments, sheets or clumps together forming a 
supercluster-void network which is remarkably similar to the
mass distribution in the Universe on large scales (above
$4-5$ Mpc) as shown in numerical simulations ({\it e.g.} Sahni and Shandarin 1996) and 
redshift galaxy catalogs ({\it e.g.} Bharadwaj et al. 2000). 
However, at small scales, within collapse structures, inertial approximation 
of Zel'dovich breaks down. 

The question arises whether caustics are 
relevant to the physics of dark matter at small scales,
for example at the scale of a 
dark matter halo.
Dark matter haloes can form for instance from the triaxial collapse of
spherical perturbations or accretion of matter into the overdensities at 
the junctions of the filaments. In a cold dark matter Universe (with or without
cosmological constant), they grow
in a hierarchical manner by merging with other haloes and by accreting mass and are hosts to the 
formation of galaxies. In numerical simulations, they are
bound overdense regions which are identified 
by various percolation algorithms.
High resolution simulations, 
surprisingly, have found that 
the spherically-averaged equilibrium density profiles 
of cold dark matter haloes 
can be described by a power law (Dubinski \& Carlberg 1991) 
and universal two-parameter function (Navarro, Frenk \& White 1996, 1997).
Although recent simulations have clearly
established that many of the
supposedly "relaxed" haloes still contain a large number of smaller subhaloes
(Klypin et al. 1999, Moore et al. 1999) they do not show the presence of discrete
flows and caustics (Moore 2001, Helmi, White \& Springel 2003). 
We believe that they
have not yet achieved enough mass resolution to observe the small-scale caustics in three dimensions and also probably suffer from spurious
collisional effects (Melott et al. 1997, Splinter et al. 1998, Binney 2004) which 
wash out the caustics.

It is worth remarking at this point that recent high resolution 
N-body simulations of the neutralino-dominated 
Universe have shown that a considerable  number 
of the smallest 
haloes with masses as small as $10^{-6} M_{\odot}$ survive until
the present epoch (Diemand, Moore \& Stadel 2005). These simulations  
suggest that there must be about $10^{15}$ such haloes in our galaxy.
These haloes are expected to have very smooth caustics 
because there had been no 
smaller scale fluctuations in the initial spectrum. The results
of our work are most directly applicable to this type of structures.  

Analytic evaluation of the halo
density profile, and the prediction of the existence of caustics inside these
structures, started with the works of
Gott (1975) and Gunn (1977) who used the spherically symmetric model. 
With the main objective of
explaining the flattening of the rotation curves of the galaxies,
they considered the formation of a dark matter halo
from the secondary infall of matter onto an already formed galaxy (or in
later works onto a spherical overdense region).
In an Einstein-de Sitter Universe a spherical overdensity
expands and then turns around to collapse.
After collapse and at late times, the fluid motion becomes
selfsimilar: its form
remains unchanged when length are re-scaled in terms of the radius of the 
shell which is currently turning around and falling onto the galaxy.
Physically selfsimilarity arises because gravity is
scale-free and because mass shells outside the initial overdensity are
also bound and turn around at successively later times.
Self-similar solutions give power-law density profiles
whose exact scaling properties depend on the central boundary
conditions and on whether the fluid is collisionless or collisional
(Fillmore and Goldreich 1984, Bertschinger 1985a, 1985b). The density profile
obeys a power-law on the scale of the halo which provides an explanation of the
flattening of the rotation curves of the galaxies. However, on smaller scales the
density profile contains many spikes
({\it i.e.} caustics) of infinite density (with an artificial cutoff due to
finite numerical resolution). The position and the time of formation
of these caustics are among many properties which have been studied in the
secondary infall model (Bertschinger 1985b).\footnote{Although various elaborations 
have since been made on the secondary infall, in order to accommodate for the bi-scaling of
the haloes density profile observed in simulations ({\it e.g.} see Henriksen 2004) here we 
concentrate on the original secondary infall model which yields a pure power law density profile.}

Such aforementioned studies not only
have proven valuable for the prediction and
description of large-scale structure, of dynamics and distribution of mass inside 
dark matter haloes and of galaxy formation but also recently
for the detection of dark matter particles.
Due to their
significantly high density over their often already dense background, 
and their large number density, caustics are clearly of importance
for dark matter search experiments.

In the past few years, major experiments have gone underway for the {\it
direct} and {\it indirect} detection of dark matter 
particles. Direct detection experiments, such as {\small DAMA} and {\small EDELWEISS}, 
often use the annual modulation of the
signal due to the orbital motion of the 
Earth around the Sun.
Since the flux of dark matter
in direct searches depends {\it linearly} on the local dark matter density,
the search strategy and data analysis strongly depend on the spatial 
distribution of dark matter and its dynamics in the galactic halo.

The indirect-detection experiments, such as {\small ANTARES}, {\small HESS} 
and {\small GLAST}, search for    
products of annihilation of dark matter 
candidates ({\it e.g.} neutralinos) such as energetic neutrinos and $\gamma$-rays. 
In the indirect searches, the flux of the annihilation products
depends {\it quadratically} on the local dark matter density.
Thus the degree of clumpiness, the density profile of a dark matter
halo, the presence or absence of a central
supermassive black hole and finally the presence of caustics, could influence the 
annihilation rate and boost the $\gamma$-ray flux significantly. 
It has been shown that this boost is significant if there is a cusp at
the centre of the halo (Stoehr 2003, Salati 2004). The 
accretion of dark matter into a central
black hole if present in the halo, could also boost the gamma-ray flux
by few orders of magnitude but again only if the dark matter 
profile develops a cusp at the centre
(Gondolo and Silk 1999). Thus, a central core profile would not in general lead to a
significant boost of the flux. However, although 
dark-matter-only simulations seem to show
a cuspy profile in the centre of the haloes, some of the observations seem to
contradict these predictions ({\it e.g.} see McGaugh et al. 2003).  
In addition, whether the cusp observed in the dark-matter-only 
simulations would survive in the
presence of gas or would become less steep or else disappear due to the reaction from
the baryonic gas, or is simply a numerical artifact (Binney 2004) is unclear. 
Caustics, on the other hand, would be inevitably present, as a direct
consequence of Jeans-Vlassov-Poisson equation ({\it e.g.} 
see Alard \& Colombi 2004 for a recent 
numerical  simulations in 1 dimension). Therefore, it is worthwhile to
study the density enhancement in the caustics and its possible implications
for dark matter search experiments. 
Many  properties of the dark matter halos related 
to detection of the dark matter signatures has been already discussed in early works. 
For example
the velocity magnitudes of the peaks in the velocity space and the large-scale
properties of galactic halos have been studied 
(Sikivie, Tkachev \& Wang 1997) and the 
the geometry of caustics in the galaxy halos has been discussed
(Sikivie 1998, Sikivie 1999, Sikivie \& Ipser 1998).
A simple estimation of maximum density in caustics
due to small thermal velocity dispersion has also been carried out
(Bergstr\"om, Edsj\"o \& Gunnarsson 2001). \
The formation and role of micropancakes in halos has been discussed (Hogan
2001). The more general question of the expected dimensionality of 
phase-space patterns in observations of galaxy structures has been studied 
and it has been suggested that the most prominent features will
be stable singularities (Tremain 2001).
  
The presence of a small velocity dispersion, {\it e.g} for neutralinos
which are presently the most plausible dark matter candidates,
smoothes the matter density at the caustic and gives it a finite maximum
value. The principle problem addressed in this work is 
the precise derivation of the value of this quantity 
and its implications for dark matter search experiments.

In view of the fact that almost all dark matter candidates have non-negligible
velocity dispersion,
we consider the secondary infall of dark matter with a very
small but finite velocity dispersion.
In this case the density profile at the very vicinity of the caustic would be
affected and in addition caustics would have a physical cut-off to their density. 
Here, we evaluate analytic expressions for the density profile at the
vicinity of the
caustics and also determine the maximum density at the caustic positions.
The analytic expression for the
density profile of the caustics is given as a function of the initial velocity
dispersion. The $\gamma$-ray emission measure from the annihilation 
of the neutralinos in the caustics is then evaluated. 
Using our results, we evaluate the position, the thickness, the density and 
the $\gamma$-ray emission measure for the first caustic of M31.

This article is organised as follows.
In Section \ref{sec:selfsimilar}, we review the basics of the secondary infall model.
In section \ref{sec:maximum density}, we derive analytic expressions for 
the density profiles near the caustics in the presence of
small velocity dispersion.
In Section \ref{sec:emission measure}, we use 
our density profile and evaluate a general
analytic expression for the emission measure 
from a typical caustic. In Section \ref{sec:andromeda} we 
evaluate the $\gamma$-ray emission measure from
dark matter annihilation in the first caustic (nearest to us) of M31.
In Section \ref{sec:conclude} we conclude our main results.

%!!!!!!!!!!!!!!!!!!!!!!!!!!!!!!!!!!!!!!!!!!!!!!!!!
\section{Self-similar model}
\label{sec:selfsimilar}
%!!!!!!!!!!!!!!!!!!!!!!!!!!!!!!!!!!!!!!!!!!!!!!!!!

We consider a spherical overdensity of collisionless fluid
in an Einstein-de Sitter Universe which eventually ceases expansion
and turns around to collapse. 
The trajectory of a fluid element in radial motion obeys Newton's law
\be
{d^2r\over dt^2}=-{Gm(r,t)\over r^2}\,,
\label{newton}
\ee
where the mass, $m(r,t)$ inside a radius $r$, is not constant 
due to shell-crossing. 
At first the only way to tackle this problem
seems to be via an N-body simulation. However,
a major simplification arises once it is realised that the problem has
a similarity solution [\cite{fg84} and \cite{Bert85b} and 
we use the notations of \cite{Bert85b} throughout]. 
The turn around radius $r_{\rm ta}(t)$ which is the only length scale in
the problem is used to introduce the non-dimensional variables
\be
\lambda = {r(t)\over r_{\rm ta}(t)}\quad;\quad
\xi={\rm ln}\left({t\over t_{\rm ta}}\right)\quad;\quad
M(\lambda)={3\over 4\pi}{m(r,t)\over \rho_H r_{\rm ta}^3}\,,
%m(r,t)={4\pi\over 3}\rho_H r_{\rm ta}^3 M(\lambda)\,,
\label{nondim1}
\ee
where
\be
r_{\rm ta}(t) = r_{\rm ita}\left({t\over t_{\rm ita}}\right)^{8/9},
\label{eq:rta}
\ee
$r_{\rm ita}$ is the initial turnaround radius, $t_{\rm ita}$ is
the initial turnaround time, $\rho_H$ is the
Einstein-de Sitter density ($\rho_H=1/6\pi Gt^2$) 
and $t_{\rm ta}$ is the turnaround time for
a given particle ({\it i.e.} when the particle is at its largest radius). 
In terms of the nondimensional variables (\ref{nondim1})
Newton's equation (\ref{newton}) becomes
\be
{d^2\lambda\over d\xi ^2}+{7\over 9}{d\lambda\over d\xi}
-{8\over 81}\lambda=-{2\over 9\lambda^2} M(\lambda)\,,
\label{newtonnondimensional}
\ee
which has no explicit dependence on non-dimensional time $\xi$. 
The equation should be solved with the initial condition (at $\xi=0$
corresponding to $t=t_{\rm ta}$)
\be
\lambda=1; \qquad {d\lambda\over d\xi}=-{8\over 9}\,,
\label{initialcondition}
\ee
and a prior knowledge of the mass $M(\lambda)$. 
In the case of the Hubble flow,
there is a simple solution to equation 
(\ref{newtonnondimensional}), before shell-crossing, which is given by
$M(\lambda)=\lambda^3=M_{\rm ta} e^{-2\xi/ 3}$.
However, after shell-crossing occurs 
there are many particles having the same value of
$\lambda$. This can be taken into account simply by the summation
\be
M(\lambda)=M_{\rm ta}\sum_i (-1)^{i-1}e^{-2\xi_i/ 3}
\label{masseqn}
\ee
which adds (for $i$ odd) the mass of the particles interior to $\lambda$ and 
subtracts (for even $i$) the mass exterior to $\lambda$, accounting 
correctly for shell-crossing. 
Equations (\ref{newtonnondimensional}) and
(\ref{masseqn}) can be solved numerically by iteration (see Appendix \ref {appendix:Numerical simulations}
for more detailed account).
Here, we take a simpler approach. At small values
of $\lambda$ ($\lambda \ll 1$), mass becomes a 
power-law $M(\lambda)\approx 11.2 \lambda^{3/4}$ (Bertschinger 1985b). 
We take this fact into account and instead of 
solving (\ref{newtonnondimensional}) and (\ref{masseqn}) iteratively, use
a simple approximation for $M(\lambda)$
\be
M(\lambda)\approx{11.2\lambda^{3/4}\over 1+\lambda^{3/4}}.
\label{massfit}
\ee
and then solve (\ref{newtonnondimensional}) numerically at the given 
gravitational potential generated by the mass distribution (\ref{massfit}). 

As shown in Appendix \ref{appendix:Numerical simulations} by 
Figure \ref{fig:massfig} the approximation (\ref{massfit}) generates relatively small errors.
A notable discrepancy between the approximation (\ref{massfit}) and expression
(\ref{masseqn}) appears only at the relatively large values 
of $\lambda\sim 1$. However, for all
the caustics under consideration the value of $\lambda$ is far less than one
[the largest value of $\lambda$ for 
the first caustic is at $\lambda\approx 0.36$ (see the table in Fig. \ref{fig:massfig})].

Solutions to equations (\ref{newtonnondimensional}) and (\ref{masseqn}), give a
power-law density profile convolved with many sharp spikes (the caustics). As a particle
expands to its turnaround radius, it collapses 
and re-expands again to its new maximum radius, which
gives the time and position of the first caustics. It 
then re-collapses and re-expands to the position of 
the second caustic and so on (see Fig. \ref{fig:caustics}).
The calculation of the halo density profile itself is not 
the subject of this work. Here we
are primarily concerned with the calculation of the density profile near the caustics
and the maximum density at the caustics in 
the case of dark matter with finite velocity dispersion.

%%%%%%%%%%%%%%%%%%%%%%%%%%%%%%%%%%%%%%%%%%%%%
\section{Maximum density in cosmological caustics}
\label{sec:maximum density}
\subsection{Cold medium}
First, we derive the equations of motion in terms of physical time
$t$ and radius $r$. From the definitions (\ref{nondim1}) one 
can easily obtain
\ba
t&=&t_{\rm ita}\left(\frac{R_{\rm ta}}{r_{\rm ita}}\right)^{9/8} e\,^{\xi} ,
\label{eq:t_xi}\\
r&=& R_{\rm ta}\, \exp\left({\frac{8}{9}\xi}\right) \lambda(\xi),
\label{eq:r_xi}
\ea
where 
\be
R_{\rm ta}\equiv r_{\rm ta}(t_{\rm ta})
=r_{\rm ita}\left(\frac{t_{\rm ta}}{t_{\rm ita}}\right)^{8/9}
\label{eq:R_ta}
\ee
is the turnaround radius reached by a particle at 
its turnaround time (which is equivalent to $r_{\rm ta}^\prime$ in Bertschinger 1985b).
Solving equation (\ref{eq:t_xi}) for $\xi$ one can obtain 
in terms of the physical coordinates and function $\lambda=\lambda(\xi)$
the explicit solution
\be
r(R_{\rm ta},t)=r_{\rm ita}\left(\frac{t}{t_{\rm ita}} \right)^{8/9}
 \lambda\left\{\ln \left[\frac{t}{t_{\rm ita}}\left( \frac{R_{\rm ta}}{r_{\rm ita}}\right)^{-9/8}\right]\right\}.
\label{eq:r_R_ta} 
\ee
Introducing dimensionless time $\tau$ and dimensionless coordinates 
$x$ and $q$ 
\be
\tau=\frac{t}{t_{\rm ita}},\quad x=\frac{r}{r_{\rm ita}}, \quad
\mbox{and} \quad q=\frac{R_{\rm ta}}{r_{\rm ita}}
\label{eq:dimlesst}
\ee
one can further simplify equation (\ref{eq:r_R_ta})
\be
x=\tau^{8/9} \lambda[\ln(\tau q^{-9/8})].
\label{eq:x_tau}
\ee
Equations (\ref{eq:r_R_ta}) and (\ref{eq:x_tau}) represent the mapping 
from Lagrangian space
to Eulerian space parameterised by time.

The density can be obtained from the conservation of mass
\be
\rho(x) =\rho(q) \frac{q^2}{x^2} \left|\frac{d x}{d q}\right|^{-1},
\label{eq:mass_conserv}
\ee
where the ratio $d x/d q$ must be taken at the time 
of formation of the caustic, $\tau_k$.
The condition of the caustic formation $d x/d q =0$
requires 
\be
\lambda^{\prime}(\xi_k)=0
\label{eq:caust_cond}
\ee
where $\lambda^\prime = d\lambda/d\xi$ 
and $\xi_k =\tau _k q^{-9/8}$ denotes the position of a  
maximum of the function $\lambda(\xi)$. The derivative 
$d x/d q$ at the Lagrangian distance $\Delta q$ from the caustic is
\be
\frac{d x}{d q}=\left(\frac{\partial^2 x}{\partial q^2}\right)_{\tau_k} 
\Delta q
= \frac{81}{64} \frac{\tau_k^{8/9}}{q^2}\lambda^{\prime \prime}_k\Delta q\,,
\ee
where $\lambda^{\prime\prime}_k = \lambda^{\prime \prime}(\xi_k)$
The relation between $\Delta x$ and $\Delta q$ can be easily found by
expanding  $x(\tau,q)$ [equation (\ref{eq:x_tau})] into the Taylor series
at the time $\tau_k$ and using the condition of $\lambda=max$ 
[given by equation (\ref{eq:caust_cond})]
\be
\Delta x = \left(\frac{\partial x}{\partial q}\right)_{\tau_k} \Delta q 
+\frac{1}{2}\left(\frac{\partial ^2 x}{\partial q^2}\right)_{\tau_k} \Delta q^2
=\frac{1}{2}
\frac{81}{64} \frac{\tau_k^{8/9}}{q^2}\lambda^{\prime \prime}_k \Delta q^2.
\ee
Thus, the inverse derivative $(dx/dq)^{-1}$ becomes 
\be
\left|\frac{d x}{d q} \right|^{-1}
= \frac{1}{2}\frac{8}{9}\frac{q}{\tau_k^{4/9}}
\left(-\frac{\lambda^{\prime \prime}_k}{2}\right)^{-1/2}
(-\Delta x)^{-1/2},
\ee  
where the signs in the above equation reflect the signs of 
$\lambda^{\prime\prime}_k <0$ and $\Delta x <0$ in the vicinity of the caustic.
Substituting the derivative $(dx/dq)^{-1}$, obtained above, into
equation (\ref{eq:mass_conserv}) 
one derives the density in the vicinity of a caustic
(additional factor 2 must be added due to two stream flow at $\Delta x <0$)
\be
\rho(\Delta x) = A_k \left(-\Delta x \right)^{-1/2},
\label{eq:rho}
\ee
with
\be
A_k=\frac{2}{9} \rho_H \frac{M_{ta}}{\lambda_k^2}
 \exp\left(-\frac{2}{3}\xi_k \right) 
\left(-\frac{\lambda^{\prime\prime}_k}{2} \right)^{-1/2} x_{ta}^{1/2},
\label{eq:A_k}
\ee
where $\rho_H=1/6\pi Gt^2$ is the mean density of the universe, 
$M_{ta}=(3\pi /8)^2$, $\lambda_k=\lambda(\xi_k)$, and 
$x_{ta}=r_{ta}(t)/r_{ita}$
is the present dimensionless turnaround radius.
[Equation (\ref{eq:rho}) corresponds to equation (4.7) in Bertschinger (1985b).]
Substituting $r$ for $x$ [equation (\ref{eq:dimlesst})] 
one can obtain the density
in terms of dimensional physical parameters. 
The parameters of the self-similar solution
[$\xi_k, \lambda(\xi_k)$ and $\lambda^{\prime\prime}(\xi_k)$]
that determine the density in the vicinity of every caustic  
must be obtained from numerical integration of equation (\ref{newtonnondimensional})
and equation (\ref{masseqn}) [or (\ref{newtonnondimensional}) and
the approximate equation (\ref{massfit})].

%%%---------------------------------------------------
\subsection{Medium with thermal velocity dispersion}
\label{sec:Med_with_Therm_Vel}
%%%---------------------------------------------------

In this section, we derive semi-analytic expressions for 
the density profile in the vicinity of the caustics for non-zero velocity
dispersion following the method used in Zel'dovich and Shandarin (1982)
and Kotok and Shandarin (1987). 
Although in rigorous mathematical terms caustics, defined as
manifolds of infinite density, would not form in the presence of 
a finite velocity
dispersion, the density at the caustic ``positions'' would still be 
extremely high if the velocity dispersion 
is very small, as is the case for most dark matter candidates,
and hence we still refer to these sharply-dense structures as caustics. 

The motion of a  medium with small thermal velocity dispersion can be 
approximated as a simultaneous evolution of many streams with different
initial velocities $v$ at $\tau=\tau_{\rm ta}$. The formation of the caustic in 
every stream occurs at different radii $x_v$.
We denote the distance from the caustic as $\delta x_v=x-x_v$. 
We assume a linear relation 
between the relative position of the caustic and the initial velocity
of the stream $v$ 
\be
x_v -x_0=\alpha_k v, \label{eq:caustic_dist}
\ee
where $v$ is the dimensionless velocity related to physical velocity
$u$ as $u= (r_{ita}/t_{ita})v$ and
$\alpha_k$
is a negative constant to be determined numerically for every caustic.
We will express the density as a function of the distance
$\Delta x = x-x_0$ from the caustic in the stream with zero initial velocity
\be
\delta x_v= \Delta x -\alpha_k v.
\label{eq:alpha}
\ee
The major effect to consider in determining the maximum density
in the `caustic' in the medium with small thermal velocities 
is the shift of the position of the caustic
in every stream with respect to the stream with $v=0$. 
As a  result the $1/\sqrt{-\Delta x}$ factor in equation (\ref{eq:rho})
must be modified as in the following integral 
\be
\rho(\Delta x)=\int \frac{f(v)dv}{\sqrt{-\Delta x+\alpha_k v}},
\ee
where $f(v)$ is the velocity distribution function at the turn around
radius $q$ and the turn around time $\tau_{\rm ta}$ corresponding to a chosen
caustic and $f(v)$ is normalised to unity: $\int f(v) dv =1$.
The above integral is simply the sum of densities in all streams at
a distance $\Delta x$ from the true caustic in cold medium.

In the simplest case of the one-dimensional top-hat (TH) velocity distribution
\be
f_{TH}(v)=\cases{ \frac{1}{2\sigma_v}, \qquad
\qquad {\rm if }\quad|v| < \sigma_v \cr  0\qquad\qquad\quad {\rm otherwise} \cr}
\ee
we obtain
\be
\rho(\Delta x) = \frac{A_k}{\sqrt{|\alpha_k\sigma_v|}} \cases{
\sqrt{1-\frac{\Delta x}{|\alpha_k \sigma_v|}}
-
\sqrt{-1 -\frac{\Delta x}{|\alpha_k\sigma_v|}}
\qquad\qquad \ \  \Delta x \le -|\alpha_k\sigma_v|, \cr \cr
\\
\sqrt{1-\frac{\Delta x}{|\alpha_k\sigma_v|}}
\qquad \qquad \qquad\qquad   \ \ 
-|\alpha_k\sigma_v| \le \Delta x \le |\alpha_k\sigma_v|, \cr \cr
\\
0
\qquad \qquad \qquad \qquad\qquad   \ \ \Delta x \ge |\alpha_k\sigma_v|,
\cr}
\label{eq:den_TH}
\ee
for the density profile near caustics, where $A_k$ is given by equation (\ref{eq:A_k}).
One recovers equation (\ref{eq:rho}) from equation (\ref{eq:den_TH}) in the limit of
$\sigma_v =0$.

In order to find the constant $\alpha_k$ we numerically solve the equation
for $\lambda(\xi)$ in the presence of the initial nondimensional velocity
perturbation $\delta \lambda^{\prime}_0$ in the initial conditions at 
$\xi=0$ so that
\be
\lambda(0)=1, \qquad \lambda^{\prime}(0)
=-\frac{8}{9}+\delta\lambda^{\prime}_0.
\label{eq:pert_init_cond}
\ee
The change of the maximum of 
$\lambda(\xi_k) = max = \lambda_k \rightarrow \lambda_k+\delta \lambda_k$
in this case is found to be well approximated by a linear 
expression
\be
\delta \lambda_k=\Lambda_{k} \delta\lambda^{\prime}_0,
\label{eq:delta_lambda_k}
\ee
where the coefficient $\Lambda_{k}$ depends on the caustic $k$
and shall be determined numerically.
Differentiating equation (\ref{eq:x_tau}) one finds the relation 
\be
\frac{dx}{d\tau}=\tau^{-1/9}\left[\frac{8}{9}\lambda(\xi)
+\lambda^{\prime}(\xi)
\right],
\ee
between the dimensionless velocity $dx/d\tau$ and nondimensional functions.
Therefore at the turnaround time, $\tau_{\rm ta}$, the dimensionless velocity, $v$, 
given by equation (\ref{eq:caustic_dist}), is
\be
v=\tau_{\rm ta}^{-1/9}\delta\lambda^{\prime}_0.
\ee
Taking the variation of equation (\ref{eq:x_tau}) at $\tau_k$ and using
equation (\ref{eq:delta_lambda_k}) one obtains the distance between the caustic
in the stream with velocity $v$ and the caustic in the stream with $v=0$
\be
x_v-x_0 = \tau_k^{8/9} \Lambda_{k}\delta\lambda^{\prime}_0. 
\ee
Finally, recalling that $x_v-x_0 =\alpha_k v$ [equation (\ref{eq:caustic_dist})] 
one obtains
the expression for $\alpha_k$ in terms of $\Lambda_k$
\be
\alpha_k=\tau_{\rm ta}\exp\left(\frac{8}{9}\xi_k\right)\Lambda_{k},
\label{eq:alpha_k}
\ee
noting that $\Lambda_{k} <0$ and $\alpha_k <0$.   
Combining equations (\ref{eq:A_k}),(\ref{eq:den_TH}) and (\ref{eq:alpha_k}) gives 
the full expression for the density in the vicinity of caustics
for the top-hat velocity distribution function.
We derive the density profile in the vicinity of a caustic
for the exponential and Gaussian velocity distribution functions in
Appendix \ref{appendix:exp_gauss}.

\begin{figure}
\bc
\includegraphics[width=107mm]{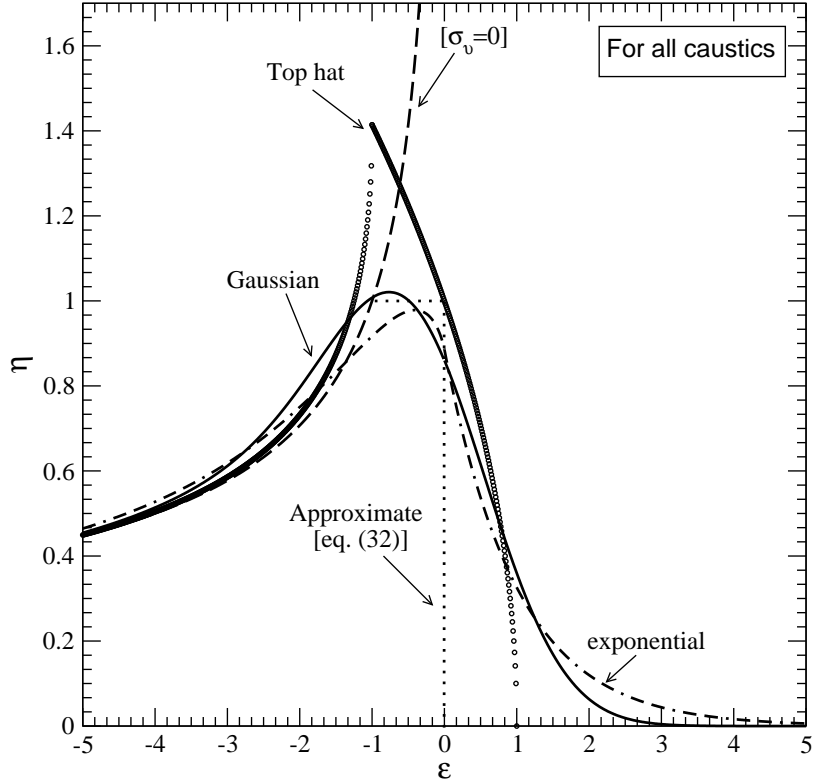}
\ec
\caption{The rescaled density 
$\eta=\rho A_k^{-1} \left(\alpha_k \sigma_v\right)^{1/2}$  versus the rescaled distance 
$\epsilon=\Delta x \alpha_k^{-1} \sigma_v ^{-1}$ 
from the caustic. Dashed line shows the case of cold matter ($\sigma_v=0$),
circles show the top-hat velocity distribution. Dashed-dotted
and solid lines show the cases of the exponential and the Gaussian initial velocity
distributions respectively. 
The dotted curve is a simple approximation 
given by equation (\ref{eq:den_appr}), which is used 
in the evaluation of the emission measure in 
Section \ref{sec:emission measure} and used also in Figure \ref{fig:detection}.
The density profile is evidently universal: it is independent of
halo parameters and is valid for all of the caustics.
}
\label{fig:den_prof}
\end{figure}

A much simpler and approximate expression for the density can be written as
\be
\rho(\Delta x)
=A_k
\cases{ 
\left(-\Delta x\right)^{-1/2} 
\quad\quad \ \ \mathrm{for}\ \quad 
\Delta x \le \Delta x_{ck},  \cr\cr
 \left(-\Delta x_{ck}\right)^{-1/2}\ \ \qquad \mathrm{for}\ 
\Delta x_{ck} \le \Delta x \le 0,     \cr\cr
0 \ \ \qquad\qquad\qquad\quad \mathrm{for} \ \  \Delta x > 0.
\cr}
\label{eq:den_appr}
\ee
where 
\be
\Delta x_{ck}=\alpha_k\sigma_v=\sigma_v \tau_{\rm ta} \exp\left(\frac{8}{9}\xi_k\right) \Lambda_k,
\label{eq:delta_xc}
\ee
note that $\Delta x_{ck}$ is negative.
The above expression approximates the maximum density for the exponential and Gaussian
velocity distribution functions quite well (see Fig. \ref{fig:den_prof}).
Physically this 
approximation means that the highest density in caustics is reduced by
thermal velocities but since the amount of mass having the highest densities
is small $\propto (-\Delta x_{ck})^{1/2} \propto \sigma_v^{1/2}$ it does not
significantly affect the rest of the density distribution.  Although a
similar idea was used in the estimate by Bergstr\"om, Edsj\"o, Gunnarsson (2001)
they have not derived the thickness of the caustic (eq. \ref{eq:delta_xc} and
\ref{eq:delta_rk})
and have not shown the accuracy of their approximation. 

In order to evaluate the density profile in the vicinity of a caustic we
numerically solve equation (\ref{newtonnondimensional}) 
with the mass distribution approximated by equation (\ref{massfit}) 
and the initial conditions given by equation (\ref{eq:pert_init_cond}).
We summarise the caustic parameters (including $\Lambda_k$) in Table 1. 

\centerline
{
\brr[t]{|c|c|c|c|c|}
\hline
\quad  k  \quad & \xi_k & \lambda_k=\lambda(\xi_k) & \lambda^{\prime \prime}_k=
\lambda^{\prime \prime}(\xi_k) &\quad \Lambda_k \quad\\
\hline
\hline 
1  &  0.985 &    0.368  &  -5.68  &  -0.070   \\
2  &  1.46  &    0.237  &  -11.2  &  -0.025   \\  
3  &  1.76  &    0.179  &  -16.7  &  -0.014   \\ 
4  &  1.98  &    0.146  &  -22.3  &  -0.0085  \\
5  &  2.16  &    0.124  &  -28.0  &  -0.0059  \\
6  &  2.31  &    0.108  &  -33.9  &  -0.0043  \\ 
7  &  2.43  &    0.0960 &  -39.8  &  -0.0034  \\
8  &  2.55  &    0.0866 &  -45.7  &  -0.0027  \\
9  &  2.64  &    0.0790 &  -51.7  &  -0.0022  \\
10 &  2.73  &   0.0728  &  -58.8  &  -0.0018  \\
\hline
\err
\label{table:causparam}
}
%-----------------------------------------------------------------------------------------
\vspace*{0.2cm}
\noindent
{\small 
Table 1. The nondimensional 
parameters of the first ten caustics obtained
 from the numerical fits to the curves in the phase space (similar to
  those shown in Fig. \ref{vlambdapeaks-20march2004-flipped})
  are shown.
}
%----------------------------------------------------------------------

Rearranging (\ref{eq:den_appr}), we arrive at
the following full expression for 
the maximum density (which occurs at $\Delta x=\Delta x_{ck}$)
\be
\rho_{\rm max}={\sqrt 2\pi^2\over 8}
{e^{-11\xi_k/18}\over \sqrt{\lambda_k^{\prime\prime}\Lambda_k}}
{1\over \lambda_k^2}\,\, \left(t\over t_{\rm ita}\right)^{-1/18}
{\rho_H\over \sqrt{\sigma_v}},
\ee
where $t$ is the present time.
Factor $(t/t_{\rm ita})^{-1/18} \sigma_v^{-1/2}$
can be expressed in terms of radial component of the physical thermal
velocity dispersion, $\sigma_{\rm ph}$, at the 
turnaround radius $R_{ta}$ at the turnaround
time $t_{\rm ta}$. First, we note that 
$\sigma_v=\sigma_{\rm ph}(r_{\rm ita}/t_{\rm ita})^{-1}$, and
then we use equation (\ref{eq:rta}) 
to obtain $r_{\rm ita}/t_{\rm ita}=(r_{\rm ta}/t)(t/t_{\rm ita})^{1/9}$. 
Combining both
factors one obtains 
\be
\left(t\over t_{\rm ita}\right)^{-1/18} \sigma_v^{-1/2}=
\sigma_{\rm ph}^{-1/2} \left( \frac{r_{\rm ta}}{t}\right)^{1/2}.
\label{eq:sigma_ph_nd}
\ee
We wish to stress that both $t$ and $r_{\rm ta}$ are present time and
present turnaround radius while $\sigma_{\rm ph}$ is the physical
velocity dispersion at the turnaround radius at the turnaround time.
The thermal velocity dispersion at turnaround radius at the turnaround 
time can be estimated from
the conservation of the phase space volume. The density at the turnaround
radius is $D(1)=(3\pi/8)^2 \approx 1.39$ times greater than 
$\rho_H(t_{\rm ta})$ therefore $\sigma_{\rm ph}$ approximately 
$D(1)^{1/3} \approx
1.24$ times greater then the thermal velocity dispersion in the homogeneous
universe at that time.
%----------------------------------------------------------------------------------- 
%---------------------------------------------------------------------------- 
%Clearly, the ratio $(t_{\rm ita}/t_{\rm now})^{1/18}$ has 
%a value very near unity.
Thus, the maximum density only depends on the caustic, the velocity dispersion and
the background Einstein-de Sitter density, $\rho_H$. The ratio of the maximum density
to the background (Einstein-de Sitter density which 
should not be confused with the local halo density) is thus almost
{\it independent of any physical parameters}. 
Using caustic parameters given in the previous table, we can evaluate the
maximum density at caustic positions.
In Table 2, we
summarise the value of $\rho/\rho_H/\sqrt{\sigma_v}$ and also of the local 
halo density for the first ten caustics.

\centerline
{
\brr[t]{|c|c|c|c|c|c|c|c|c|c|c|}
\hline 
k \quad & 1  & 2 & 3 & 4 & 5 & 6 & 7 & 8 & 9 & 10 \\
\hline
\rho_{\rm max}/\rho_H/\sqrt\sigma_v &
11.  &  24. & 39.  & 56. & 74. & 95. & 117. & 139. & 165. & 190. \\
\hline
\rho_{\rm halo}/\rho_H &
12.  & 40. & 83. & 139. & 210. & 297. & 397. & 506. & 641. & 777. \\
\hline
\err
\label{table:maxdensity}
}
%-----------------------------------------------------------------------------------------
\vspace*{0.2cm}
\noindent
{\small 
{\bf Table 2.} This table gives the maximum caustic density 
evaluated using (\ref{eq:den_appr})
at the first ten caustic and also the halo density evaluated 
using the approximate expression (\ref{massfit}) at the position of the caustics. Both of
these densities are given as a ratio to the background Einstein de-Sitter
density, $\rho_H$. The non-dimensional velocity dispersion, $\sigma_v$, is
given by expression (\ref{eq:sigma_ph_nd}).
}
%------------------------------------------------------------------------------

Thus, we can evaluate for a given velocity dispersion, the positions when 
the maximum caustic density becomes equivalent to the background density.
Clearly for small values of $\sigma_v$ this would occur only at small radii for
inner caustics and vice versa . The values of velocity dispersion for cold dark matter
is very small and would be expected to be much smaller than unity. 
Thus, it is clear from Table 2 that the enhancement factor 
can be extremely high, for low values of the velocity dispersion $\sigma_v$.

To summarise this section, we also write the approximate maximum density 
of the caustics (see the ``Approximate'' profile of Fig. \ref{fig:den_prof})
and their thickness obtained by our method. We have for the maximum density of
caustics
\be
\rho_{\rm max}=
\left({\pi^{5/3}\over 2\sqrt {2\,}3^{1/3}}{e^{-17\xi_k/18}\over 
\sqrt{\Lambda_k\lambda^{''}_k\,}}{1\over \lambda_k^2}\right)
\sqrt{r_{\rm ta}(t)\over t\,\bar{\sigma}_{\rm ph}(t)}\bar\rho_H(t).
\ee
The radius of the caustic shell and its thickness  in physical units are
\be
r_k=\lambda_k r_{\rm ta}(t), \quad \Delta r_k= {(3\pi)^{2/3}\over 4}e^{5\xi_k/9}\Lambda_k\, t\, \bar{\sigma}_{\rm ph}(t)
\label{eq:delta_rk}
\ee
It is remarkable that the thickness of the caustics is universal and depends
only on the present day DM velocity dispersion and not on the mass
 of the DM halo. 

 \begin{figure}
\bc
\includegraphics[width=107mm]{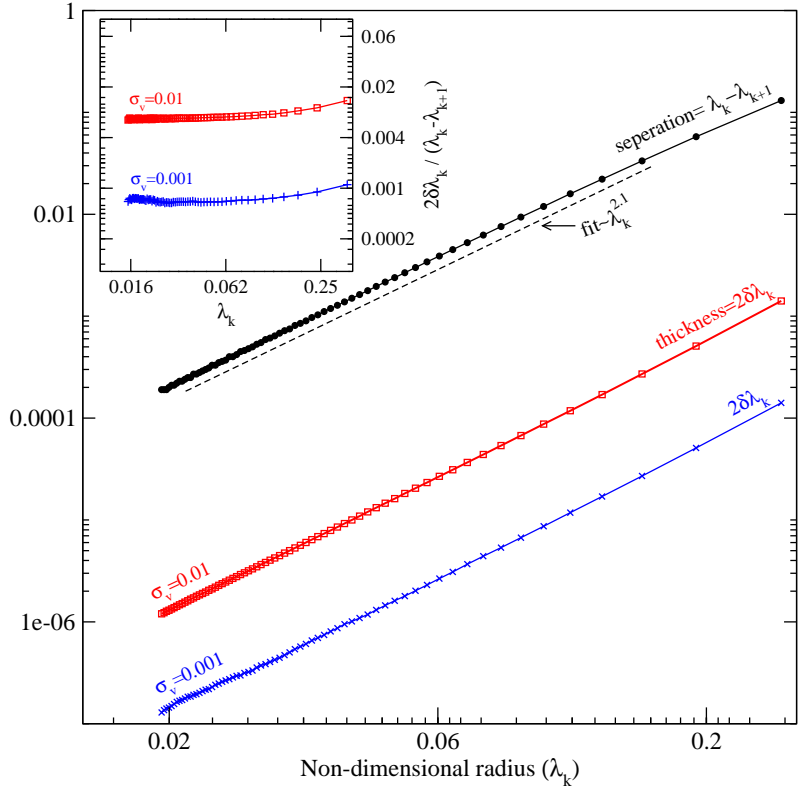}
\ec
\caption{
{\it Main plot:} The top line (filled circles) shows the separation of caustics, 
$\lambda_k-\lambda_{k+1}$, as a function of the distance, $\lambda_k$, 
from the center. 
The two bottom lines
show the distance between caustics in the streams with negative and positive
perturbations of the  initial velocity ($\delta \lambda_0^{\prime}$ in
eq. \ref{eq:pert_init_cond}). Note the logarithmic axes.
The two lower lines correspond to $\delta \lambda_0^{\prime} = \pm 0.001$ (crosses)
and $\delta \lambda_0^{\prime} = \pm 0.01$ (open squares). 
All three lines can be well fitted by a power law $\propto \lambda_k^{2.1}$
showed by the dashed line.
{\it Inset:} The ratio of the thicknesses of the caustics to their separations 
, $2\delta\lambda_k/(\lambda_k-\lambda_{k+1})$, is shown as a function of 
their radii. The vertical axis is linear.  The ratios converge to approximately
0.0007 and 0.0066 for the two different values of velocity dispersions as
marked on the plots, 
which shows its linear dependence on the initial velocity perturbation.
}
\label{fig:caustic-thickness}
\end{figure}
%----------------------------------------------------------------------------
Both the absolute  and relative  thickness 
($\Delta r_k$ and $\Delta r_k/r_k$)  of the caustic
monotonically decrease toward the center of the halo. This indicates that the radial
component of the thermal velocity decreases toward the center. We do not consider
the evolution of the angular components of the thermal velocity in this paper,
however we would like to speculate that they grow toward the center making the velocity
distribution function anisotropic (oblate ellipsoid). The decrease of the radial 
temperature can also be viewed as a consequence of the Liouville 
theorem that forbids the overlapping of the streams in the phase space.

Figure \ref{fig:caustic-thickness} shows the logarithm of separation of the caustics, 
$\lambda_k-\lambda_{k+1}$, and 
thicknesses, $2\delta\lambda_k$, [see eq.~(\ref{eq:delta_lambda_k})] and the
ratio of these two quantities (plotted in the inset)
as a function of the logarithm of the radius, $\lambda_k$.
Both the separation and thickness are scaled as a power law 
$\propto \lambda_k^{2.1}$: the formal fit to the separation of caustics is
$ \lambda_k-\lambda_{k+1} = 1.24 \lambda_k^{2.1}$ and for the two examples in the
main plot (corresponding to the initial velocity perturbations 
$\delta \lambda_0^{\prime} = \pm 0.01$ and $\pm 0.001$) this fit is re-adjusted by
factors of 0.066 and 0.007 respectively (see the inset). {\it This is a remarkable result
which shows that in the course of gravitational evolution the streams remain well-isolated
from each other in spite of the fact that their separations diminish.} 

\section{Emission measure $({\cal EM})$ from 
dark matter particle annihilation in the caustics}
\label{sec:emission measure}
%---------------------------------------------------------------------------

The annihilation flux (in {\rm photons/cm$^2$/s}) can be written as
\be
\Phi_\gamma(\psi)={N_\gamma\langle\sigma v\rangle\over 4\pi m_\chi^2}
\times {1\over \Delta\Omega}\int_{\Delta\Omega} d\Omega\,\,
\times ({\cal EM})
\label{eq:flux}
\ee
where the emission measure, 
\be
{\cal EM}= \int_{\rm line\,of\,sight}\rho^2(s) ds,
\label{eq:EM}
\ee
is found by integrating the square of the density along the line of sight and over
the solid
angle $\Delta\Omega$, $m_\gamma$ is the mass of the candidate 
particle ({\it e.g.} neutralino), $N_\gamma$ is the number 
of photons produced per annihilation. To compute the first part of integral 
$N_\gamma\langle\sigma v\rangle/(4\pi m_\chi^2)$ a supersymmetric model needs
be selected. We shall not discuss 
this aspect here which is already extensively discussed
in the relevant literatures ({\it e.g.} see Jungman, Kamionkowski 
and Griest 1996 and references therein). In this article, we obtain
analytic expression for 
the emission measure (\ref{eq:EM}), using our approximate expression 
for the density (\ref{eq:den_appr}).

Here we are interested in calculating the $({\cal EM})$ from the regions close
to the caustic surfaces formed in cold matter.
Clearly, the $({\cal EM})$ is considerably higher when the line of sight
is close to the tangent to the caustic surface. We estimate the $({\cal EM})$ 
in a small vicinity of this tangent.
Although, in principle in order to obtain the emission measure (\ref{eq:EM})
one can integrate 
expressions (\ref{etaTH}) or (\ref{etaE}) or (\ref{etaG}) numerically,  
here we use our simple approximation (\ref{eq:den_appr}) for 
the density profile in the vicinity of the caustic and make analytic estimates
for the boost factor (\ref{eq:EM}). 

\begin{figure}
\includegraphics[width=164mm]{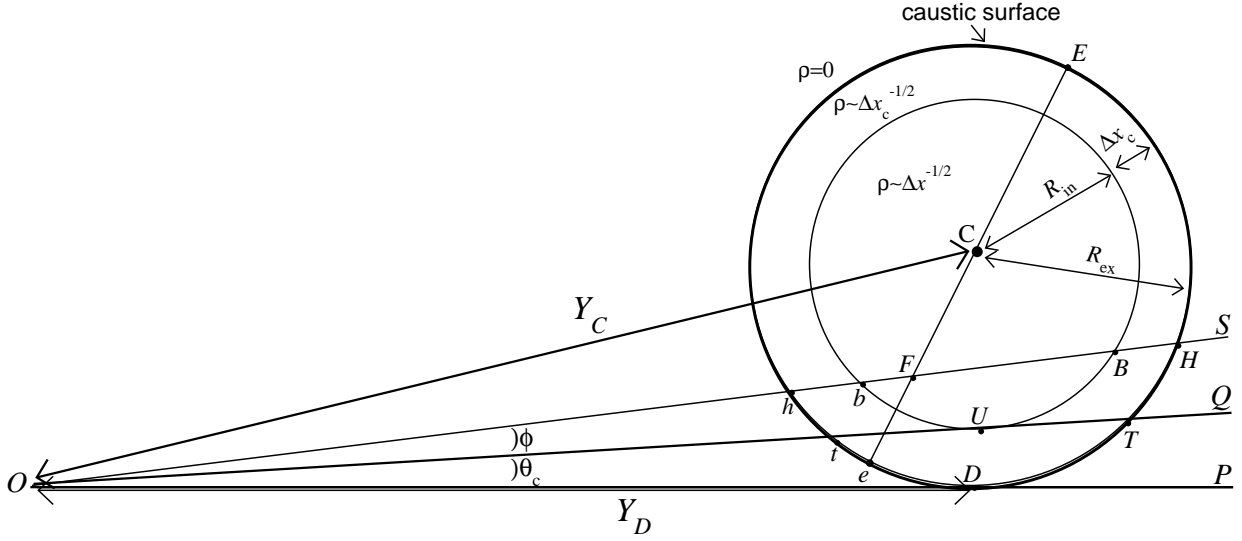}
\caption
{
An approximation is used in which the caustic is the outer-shell (larger
circle) outside which the density is zero, inside a layer of thickness 
$\Delta x_{ck}$ the density is constant taken to be $\rho_c=A_k\Delta x_{ck}^{-1/2}$. 
In the interior of this shell the density
falls as $1/\sqrt{-\Delta x}$ where $\Delta x$ is the radial coordinate  
measured from the tangent point $D$ ($\Delta x<0$). 
}
\label{fig:detection}
\end{figure}

We assume that the density in the vicinity of the caustic can be 
approximated by equation (\ref{eq:den_appr}). 
Figure \ref{fig:detection} illustrates the geometry of the system. 
The figure shows the plane passing through the observer, $O$ , the
centre of the galaxy, $C$, and point $D$ where the line of sight $OP$ is
tangential to the caustic sphere. The external and internal
caustic spheres  have the radius $R_{\rm ex}$ and 
$R_{\rm in}=R_{\rm ex}+\Delta x_{ck}$ respectively (note $\Delta x_{ck}$ is negative,
see equation (\ref{eq:deta_xc} and also Fig. \ref{fig:detection}).
The density in the shell between two spheres is constant 
$\rho = A_k (-\Delta x_{ck})^{-1/2}$,
while inside it falls as $\rho = A_k (-\Delta x)^{-1/2}$ where
$\Delta x$ is the radial coordinate measured from point $D$ 
on the external sphere ($\Delta x <0$).

We evaluate the $({\cal EM})$ as a function of the angle $\theta$ measured from 
the line $OP$ upwards.
%--------------------------------------------------------------------------
%--------------------------------------------------------------------------
First, we calculate the contribution to the emission measure along 
the line of sight which runs inside the angle POQ. 
In this case, no integral needs to be evaluated 
since the density is constant and the integral of density along the line of
sight is the density times the length of the chord between 
two points where the line of sight crosses the external circle (e.g. $tT$).

The contribution to the emission measure, 
(\ref{eq:EM}), from the lines crossing inside POQ 
is
\be
{\cal EM}(\theta)=\int \rho^2ds=\rho_c^2 L_{ch}(\theta),
\label{I_T}
\ee
where $L_{ch}(\theta)$ is the length of the chord.
The equation of the circle in the vicinity of the tangential point $D$
is to the lowest order
\be
\Delta x=-\frac{y^2}{2R_{\rm ex}} \qquad \mbox{or} \qquad y
=\pm\sqrt{-2R_{\rm ex}\Delta x} 
\label{Circle}
\ee
where $y$ is the coordinate along DP measured from point D.
For example, the length of the chord $tT$ is 
\be
L_{tT}=\sqrt{(y_T+y_t)^2+(\Delta x_T-\Delta x_t)^2} \approx y_T+y_t,
\ee
because $\Delta x$ is of the higher order in $y$ and thus the second 
term can be neglected ($y_T>0$ and $y_t<0$).
Neglecting the difference between $\Delta x_t$ and $\Delta x_T$ 
allows to approximate them
as $\Delta x_t\approx \Delta x_T \approx -L_{OD} \theta = -Y_D \theta$ 
(where $Y_D=\sqrt{Y_C^2-R_{\rm ex}^2}$ is the distance from the observer to
the caustic and $Y_C$ is the distance from the observer O to the centre C) 
assuming $\theta$ is small.
Combining this approximation with the second equation (\ref{Circle}) the 
length of the chord  can be written in a simple form
\be
L_{ch}=2 \sqrt{2R_{\rm ex} Y_D}\sqrt{\theta},
\label{eq:LargeChord}
\ee
and thus,
\be
{\cal EM}(\theta)=2 \rho_c^2 \sqrt{2R_{\rm ex} Y_D}\sqrt{\theta}
\qquad\mbox{for } \qquad  
\theta <\theta_c=\angle POQ \approx \frac{-\Delta x_{ck}}{Y_D}.
\ee
Next, we suppose that the line of sight is along OS in which case it 
crosses the region where the density falls as $(-\Delta x)^{-1/2}$. 
The integral
(\ref{eq:EM}) is now the sum of three parts:
\be
{\cal EM}_{hH}=({\cal EM})_{hb}+({\cal EM})_{BH}+({\cal EM})_{bB}=({\cal EM})_{bB}
+\rho_c^2 (L_{{hb}}+L_{BH}).
\ee
The length $L_{hb}+L_{BH} = L_{hH}-L_{bB}$ can be easily evaluated 
in a similar manner as before.
%%%%%\ba
%L_{BH}&=&x_H-x_B=
%{
%%\sqrt{\tan^2 \theta\left((2R^2-d^2)-2R\sqrt{d^2-R^2}\right)}
%%\over
%%1+\tan^2 \theta
%%}
%%\nonumber \\
%%&+&
%%{
%%\sqrt{
%%\tan^2 \theta\left[2R^2-2R/A^2-d^2\right] - 2 R\tan \theta\sqrt{d^2-R^2}}
%%\over
%%1+\tan^2 \theta
%%}
%%\ea
The length of the chord $hH$ is given by equation (\ref{eq:LargeChord}) while
for the chord $bB$ one needs to substitute 
$R_{\rm in}=R_{\rm ex}+\Delta x_{ck}$ for $R_{\rm ex}$, 
$\sqrt{Y_C^2-R_{\rm in}^2}$ for $Y_D$, and the angle
$\phi =\angle QOS$ for $\theta$. For small angles $\theta$ and small
$\Delta x_{ck}$ the change in the radius and distance to the caustic  result
 in the higher order corrections and can be neglected, yielding
\be
L_{bB}=2 \sqrt{2(R_{\rm ex}+\Delta x_{ck})\sqrt{Y_C^2-R_{\rm in}^2}}\sqrt{\phi}
\approx 2 \sqrt{2 R_{\rm ex} Y_D} \sqrt{\theta-\theta_c}.
\ee
where $\theta_c=-\Delta x_{ck}/Y_D=A_k^2/(\rho_c^2Y_D)$ and
all distances are given in units of equation (\ref{eq:x_tau}).
Thus, the $({\cal EM})$ from the parts of the line of sight with constant
density becomes
\be
({\cal EM})_{hb}+({\cal EM})_{BH} = 2 \rho_c^2 \sqrt{2R_{\rm ex} Y_D}\left(\sqrt{\theta} 
-\sqrt{\theta-\theta_c}\right) \qquad\mbox{for } \qquad
\theta >\theta_c= -\frac{\Delta x_{ck}}{Y_D}.
\ee
The complicated part of contribution to the integral (\ref{eq:EM}) comes from the line of
sight $bB$. Here, we actually need to calculate the integral (\ref{eq:EM})
where the density is no longer constant but falls as $(-\Delta x)^{-1/2}$. 
First of all
we need to express the distance $|\Delta x|=L_{eF}$ in
terms of the line of sight distance $z=L_{hF}$. 
This can be easily done by solving the intersecting chords relation
$L_{hF}\times L_{FH}=L_{eF}\times L_{FE}$ for $\Delta x$
($L_{FE}= 2R_{\rm ex}-|\Delta x|$ and $L_{FH}=L-z$ (where $L=L_{hH}$)
which yield
\be
({\cal EM})_{bB}=\rho_c^2\int\limits_{z_b}^{z_B}{\rho(z)^2}dz
=\rho_c^2\int\limits_{z_b}^{z_B} \,{dz\over {R_{\rm ex}-\sqrt{R_{\rm ex}^2-L_{bB}z+z^2}}}
\label{bB}
\ee
where $z_b$ and $z_B$ correspond to points $b$ and $B$ respectively. 
The integral (\ref{bB}) can be written in close form 
\be
I =
{R_{\rm ex}\over L} \left\{
{\rm arctanh}\left[
{2R_{\rm ex} L(L-2z)\sqrt{R_{\rm ex}^2-Lz+z^2}\over
z(L-z)(4R_{\rm ex}^2+L^2) -2R_{\rm ex}^2L^2)}\right]-
{\rm ln}\left({L-z \over z } \right) \right\}
-{\rm ln}\left(2\sqrt{R_{\rm ex}^2-Lz+z^2}-L+2z\right)
\label{I_final}
\ee
The limits in the integral are to linear order
\be
z_b=\sqrt{2R_{\rm ex} Y_D}\left(\sqrt{\theta}-\sqrt{\theta-\theta_c}\right) 
\qquad \mbox{and} \qquad 
z_B=\sqrt{2R_{\rm ex} Y_D}\left(\sqrt{\theta}+\sqrt{\theta-\theta_c}\right).
\ee
Substituting the limits into equation (\ref{I_final}) is straightforward
but results can be lengthy and complex. In addition, despite 
the exact form of the integral (\ref{I_final}) we use the accurate limits
only to the lowest order. Thus, we simplify the equation for $({\cal EM})$ by taking
the lowest order terms in the series expansion for small $\theta$
and $\theta_c$.
\be
 ({\cal EM})(\theta)=\rho_c^2\sqrt{\frac{2R_{\rm ex}}{Y_D}}\frac{1}{\sqrt{\theta}}
\ln\left(\frac{\sqrt{\theta}+\sqrt{\theta-\theta_c}}{\sqrt{\theta}-\sqrt{\theta-\theta_c}} \right).
%-2\sqrt{\frac{2Y_D}{R_{\rm ex}}}\sqrt{\theta-\theta_c}.
\ee
Thus, collecting various expressions for the emission measure together, we arrive at
\be
{\cal EM}=\rho_c^2 \left\{ \begin{array}
{r@{\quad\mbox{for }\quad}l } 
0 & \theta <0, \\
2 \sqrt{2R_{\rm ex} Y_D}\sqrt{\theta} &  0< \theta <\theta_c \\
2 \sqrt{2R_{\rm ex} Y_D}\left(\sqrt{\theta} 
-\sqrt{\theta-\theta_c}\right) +\sqrt{2R_{\rm ex}/Y_D} \, \theta^{-1/2} 
\ln\left(\frac{\sqrt{\theta}+\sqrt{\theta-\theta_c}}{\sqrt{\theta}-\sqrt{\theta-\theta_c}} \right)  &
\theta >\theta_c.
\end{array} \right. 
\label{eq:em_summary}
\ee
where once again we mention that $R_{\rm ex}$ is the radius 
of the caustic sphere, $Y_D$ is the distance
to the caustic surface, $\theta_c=-\Delta x_{ck}/Y_D=A_k^2/(\rho_c^2Y_D)$ and
all distances are given in units of equation (\ref{eq:x_tau}).

%%
%------------------------------------------------------------------
\begin{figure}
\bc
\epsfxsize= 11truecm\epsfbox{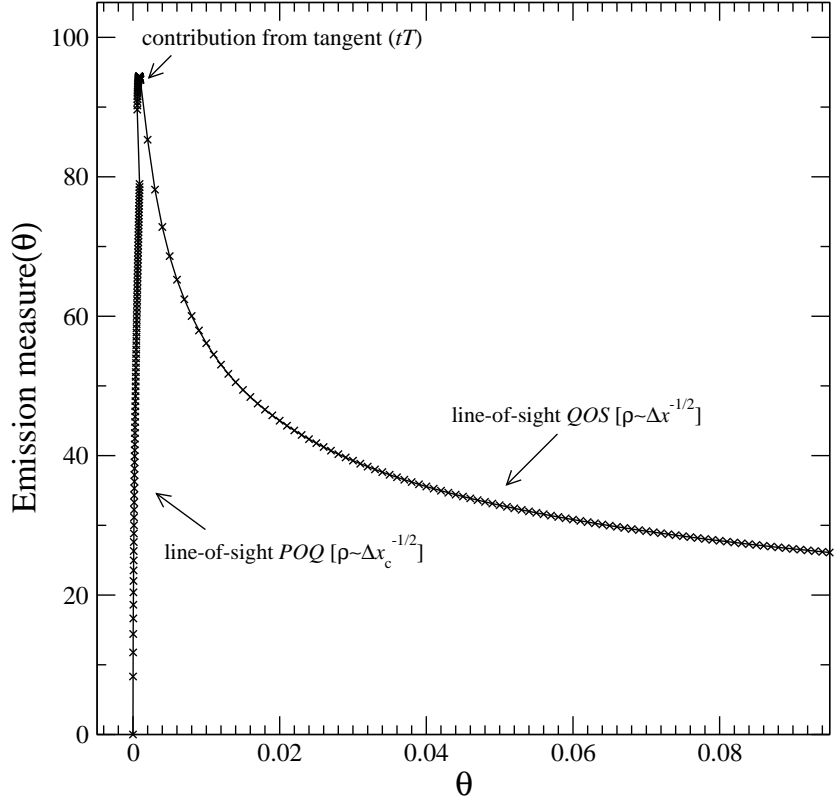}
\caption
{
Contribution to the emission measure (\ref{eq:EM}) along
different lines of sights as marked on Fig. \ref{fig:detection}. All 
values are in non-dimensional coordinates.
}
\label{fig:detectiongraph}
\ec
\end{figure}
%--------------------------------------------------------------------
%
%
%!!!!!!!!!!!!!!!!!!!!!!!!!!!!!!!!!!!!!!!!!!!!!!!!!!!!!!!!!!!!!!!
\section{Emission from a characteristic M31-type dark matter halo}
\label{sec:andromeda}
%!!!!!!!!!!!!!!!!!!!!!!!!!!!!!!!!!!!!!!!!!!!!!!!!!!!!!!!!!!!!!!!
%
%
Detailed application of our results to haloes especially minihaloes
(Diemand, Moore \& Stadel 2005) shall 
be presented in a forthcoming work.
Here, as an exercise, we apply the results 
of the previous sections to a very
simple model of M31.
Emission from M31 dark matter halo
has already been studied, largely in the context of 
dark matter search experiment 
CELEST ({\it e.g.} see Nuss et al. 2002 and Falvard et al. 2004). 
M31 is situated at a distance of 780 Kpc from the centre of MilkyWay (MW) 
with a present turnaround radius 
which is taken to be at about 800 Kpc 
(Sandage 1986, Karachentsev 2002).
In fact M31 and MW can be considered to be embedded in a
common halo, however here in our very simple model we consider
M31 to have its own halo.
The first caustic, which is the closest to us cuts the radius joining 
the centre of MW to the centre of M31 at 
a distance of about 500 Kpc from us. The tangent point to this caustic lies at 
about 700 Kpc from the centre of MW. This caustic has a thickness 
of about 0.115 Kpc and subtends an angle 
($\theta_c$) of about $0.006^o$ at the centre of Milky Way.
The maximum emission measure from 
this caustic, using the expressions (\ref{eq:em_summary})
for a velocity dispersion of $\sigma_v=0.001$ is given 
in Table 3. We assume a field of view of $1^o$, which is at the lowest-end 
for most detectors. 

\centerline
{
\brr[t]{|c|c|c|c|c|c|c|c|}
\hline
k &   r \,\, {\rm (kpc)} & \delta r\,\, {\rm (kpc)} & \theta_c^{\,o} & \rho_{\rm max} &  
\int_0^{\theta_c} {\cal EM}\, d\theta  & 
\int_{\theta_c}^{1^o} {\cal EM}\, d\theta  & {\rm Total\, 
({\cal EM})}\, {\rm Gev}^2{\rm cm}^{-5}{\rm c}^{-4} \,\\
\hline
1  &  300  &  0.115     &  0.006   &  350\rho_H   &  260\rho_H^2r_{\rm ta} 
& 5\times 10^5 \rho_H^2r_{\rm ta} &  4 \,\times\, 10^{24}\\
\hline
\err
\label{table:andromeda}
}
\vspace*{0.2cm}
\noindent
{\small 
{\bf Table 3.} This table gives {\it approximate} values for various 
caustic parameters and also
the emission measure for the first outer caustic of M31, which is the 
nearest to Milky Way.
We take the non-dimensional velocity dispersion to be $\sigma_v\sim0.001$ which 
would give only a very modest estimate of the maximum density. For neutralino,
this parameter is smaller by a few orders of magnitude leading to a 
significantly higher density maximum but at 
the same time smaller angle, $\theta_c$. Expression (\ref{eq:sigma_ph_nd}) can 
be used to transform between the
physical velocity dispersion, $\sigma_{ph}$,
and the non-dimensional velocity dispersion, $\sigma_v$.
}

The emission measure from the first caustic of M31 is 
at least of the same order\footnote{Different values for the
emission measure form the galactic centre has been 
evaluated ranging from 
$10^{21}$ to $10^{31}$ depending on various physical assumptions such as
the presence or absence of a central core, a central cusp, or 
a central supermassive black hole (see for example 
Stoehr et al. 2004; Evans, Ferrer and Sarkar 2004).} as that for example
from the centre of Milky Way.
Here, we have used a moderate 
value of $\sigma_v=0.001$, a lower value would
sharply increase the maximum density as is 
evident from expressions (\ref{eq:den_appr}) and (\ref{eq:flux}) although
it would also reduce the thickness of the caustic. The simple exercise
in this section also demonstrates that emission measure from caustics can
serve as a mean to put bound on the mass of the dark matter particle candidates.

%------------------------------------------------------------------------------
\section{Discussion}
%\section{Conclusion}
\label{sec:conclude}
In this work, we have evaluated the density profile near caustics
which arise in the selfsimilar scenario of the formation of dark matter haloes.
We have obtained a universal analytic 
expression for this density profile and its maximum value in 
the presence of a small 
velocity dispersion. We have shown that the maximum density 
at the caustics depends primarily on 
the velocity dispersion and is significantly higher than the halo density at the
position of the caustic for outer caustics. The radius at which 
the caustic density approaches the background halo density 
depends on the value of the velocity dispersion 
and is expected to arise only for the innermost caustics.
We have then evaluated the emission measure from 
the caustics and applied it to the concrete example 
of M31 type halo outermost caustic, which is closest to us. This example
demonstrates that caustics can be promising 
sources for dark matter search experiments. Application of our results 
to other haloes, in particular to small haloes which 
have not had significant mergers, to specific search experiments such as
{\small HESS} and also 
to direct detection experiments remains to be done.

Although we have considered an Einstein-de Sitter Universe, we expect
our results to give reasonable approximation for the $\Lambda$CDM Universe
as well. The role of dark energy
becomes significant at rather small redshifts ($\sim 0.3$) which
we expect to be well after the formation of the typical dark matter haloes we
consider here. Furthermore, once a particle turns around and collapses, it
separates from the background expansion and its subsequent motion should not
be effected by the $\Lambda$ term. 
However, one ought to use real density of dark matter halo
and not the critical values.
The second caveat in our
consideration is the assumption of spherical symmetry of 
haloes and their cold accretion which
does not hold in general cases. 
However, even in real 3D collapses, parts of the caustic surface can be
well-approximated by spheres.
Finally, the third and probably most 
serious problem is associated with the smooth precollapse conditions 
on galactic scale 
that contradicts the hierarchical clustering scenario. The cold dark
matter models predict a relatively high level of small scale perturbations
that result in the formation of small gravitationally bound haloes that
are assembled into more massive haloes at later times. Thus, the dark matter
accrets onto halos of galactic size in the form of smaller haloes 
that may significantly affect the density in the vicinity of caustics. 
However, there is a possibility that in cold dark matter models
the smallest haloes can survive the tidal destruction in more massive 
halos (Diemand, Moore \& Stadel 2005). In this case our results can
be applied directly. This issue will be studied in detail in the
following work.

We show that both the separation of neighboring caustics and their effective
thicknesses scale as a power law of the radius $\propto \lambda_k^{2.1}$.
This scaling demonstrates that the streams in the phase space 
corresponding to different macroscopic velocities remain well isolated
despite of the fact that the velocities of inner streams vanish at the center of the halo.
The radial component of the microscopic thermal velocities also vanishes
as the stream descends to the center. We call this effect {\it gravitational cooling}
of the radial temperature. Our method can be generalized for the case of nonradial
components of the thermal velocity and we address this question in the
following work. 

We believe that
the current results represent a step
toward building a more comprehensive theoretical model of the
gravitational collapse.   
A more elaborate study of 
caustic distributions and density profiles,
in broader settings without the assumptions of spherical symmetry, 
inertial trajectories or smooth initial conditions remains a challenging task.

\section*{Acknowledgements}

We are grateful to Joe Silk for many useful comments 
and ongoing collaboration and to Stephane Colombi, 
Pierre Salati and Brent Tully for discussions.
Special thanks go to Jacques Colin and 
Uriel Frisch for invaluable supports at
the Observatoire de la C\^ote d'Azur where major part of this work 
was carried out. 
RM was supported by a Marie
Curie HPMF-CT 2002-01532 and a European Gravitational Observatory
(EGO) fellowship at
the school of astronomy of the university
of Cardiff, UK.

%!!!!!!!!!!!!!!!!!!!!!!!!!!!!!!!!!!!!!!!!!!!!!!!!!!!!!!!!!!!!!!!!!!!!!!!!!!

%%%%%%%%%%%%%%%%%%%%%%%%%%%%%%%%%%%%%%%%%%%%%%%%%%%%%%%%%%%%%
\appendix
%%%%%%%%%%%%%%%%%%%%%%%%%%%%%%%%%%%%%%%%%%%%%%%%%%%%%%%%%%%%%%%%

\section{Numerical simulations}
\label{appendix:Numerical simulations}

There is a computational difficulty in solving equation (\ref{newtonnondimensional})
which is singular at $\lambda=0$ where 
the velocity becomes infinite and changes 
discontinuously from negative to positive.
In order to avoid this problem a small amount of angular momentum, of the form
$J^2/\lambda^3$ is added 
the RHS of equation (\ref{newtonnondimensional}) where hereafter the value
of $J=10^{-9}$ is adopted \cite{Bert85b}. 

\begin{figure*}
\bc
\epsfxsize= 16truecm\epsfbox{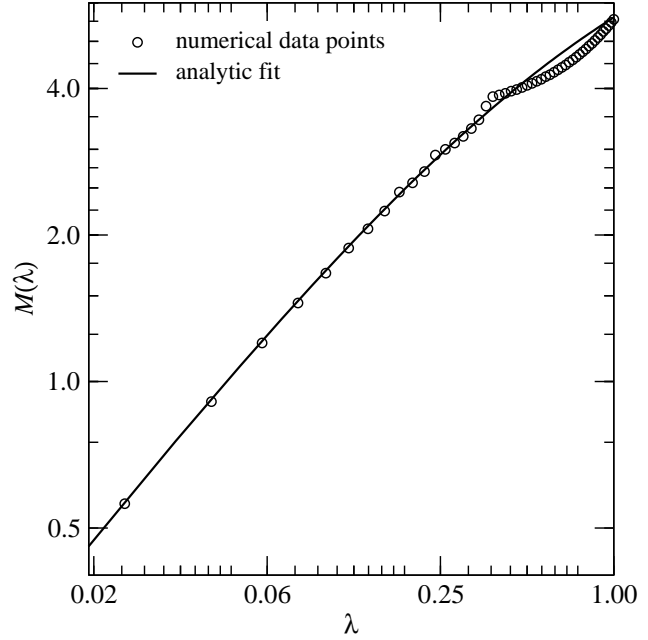} 
\ec
\caption{
The fit (solid line) $M(\lambda)={11.2\lambda^{3./4.}/ (1.+\lambda^{0.75})}$ to the
data points (open circles) taken from Table 4 of Bertschinger 1985b.  The above table
shows our values and Bertschinger 1985b values (given inside brackets) 
for $\lambda$, $\xi$ and $d^2\lambda/d\xi^2$ for the first ten caustics.
}
\label{fig:massfig}
\end{figure*}

Further computational issues arise in solving 
(\ref{newtonnondimensional}) and (\ref{masseqn}).
The outline of the numerical procedure is as follows:
a guess is first made for the mass 
distribution {\it e.g.}, $M(\lambda)=M_{\rm ta}\lambda^{3/4}$
for $\lambda\le 1$. Then equation (\ref{newtonnondimensional}) 
is integrated to obtain $\lambda(\xi)$. This is used to obtain a new 
approximation to $M(\lambda)$
and so on until a self-consistent solution is found
(see Bertschinger 1985b for full details). In this article
we have chosen the fitting formula (\ref{massfit}) instead of
(\ref{masseqn}), which as shown in Fig. \ref{fig:massfig} works 
well for small values of $\lambda$.

%----------------------------------------------------------
%---------------------------------------------------------

\begin{figure}
%\ba
%\epsfxsize= 6.7truecm\epsfbox{lambdazeta.eps} \qquad\qquad\qquad
%&&
%\epsfxsize= 6.7truecm\epsfbox{vlambdapeaks.eps}
%\nn
%\ea
\bc
\epsfxsize= 16truecm\epsfbox{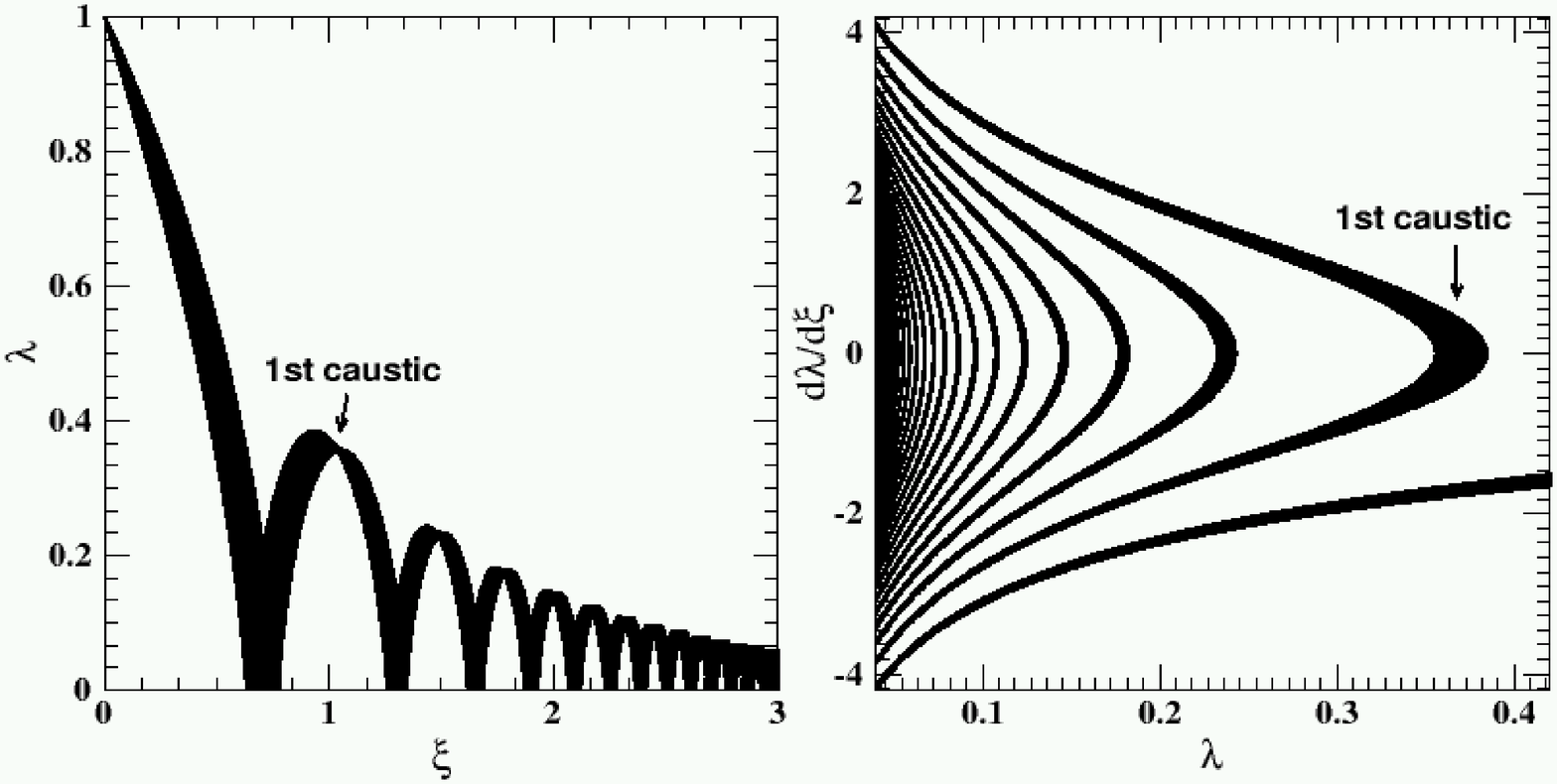}
\ec
\caption
{
The left plot is the
non-dimensional trajectory $\lambda(\tau)$ for a collisionless fluid, with
minute velocity dispersion, given by
similarity solution to equations (\ref{newtonnondimensional}) and
(\ref{massfit}). The particles reach their maximum radius at $\lambda=1$. The
non-dimensional coordinate $\lambda$ and the non-dimensional time $\xi$ at the maxima
depend almost linearly on the initial velocity of the particle.
The right plot is a small portion of the phase diagram given by the similarity
solution. Each particle travels along the entire curve and at a given time
there is a particle at each point on the curve. As we go to smaller and smaller radii,
the strips become narrower and more closely packed 
and finally {\it resemble} a smooth distribution.
To demonstrate the thickening of the caustics we have taken
an unrealistically large range of velocity dispersion, which is causing 
the shift, seen in the left panel of this figure,
in the formation times of the caustics.
}
\label{fig:caustics}
\end{figure}

The solutions to equations (\ref{newtonnondimensional}) and (\ref{massfit})
is plotted in Fig. \ref{fig:caustics} which can be viewed as the trajectory and phase
diagram of
one particle during the course of 
evolution of the halo, or as a snapshot of the positions
of many particles in the halo. The change in real 
and phase space for finite velocity dispersion is clear: velocity dispersion
leads to the broadening of caustics. The density at caustics no longer diverges
but has a maximum cut-off determined by the velocity dispersion, which is the main result
of this work.

%------------------------------------------------------------------

The small velocity dispersion effects are primarily related
to defocusing of trajectories of particles with different thermal
velocities in the vicinity of caustics. Imagine
the evolution of large number of streams each corresponding to a particular
value of thermal velocity then these streams will produce a caustic at
a slightly different radii. The resulting density field becomes the sum
of densities in every stream. We assume that each stream evolves in the
same gravitational field generated by the mass distribution of the
cold medium [equation (\ref{massfit})].

\begin{figure}
\bc
\includegraphics[width=107mm]{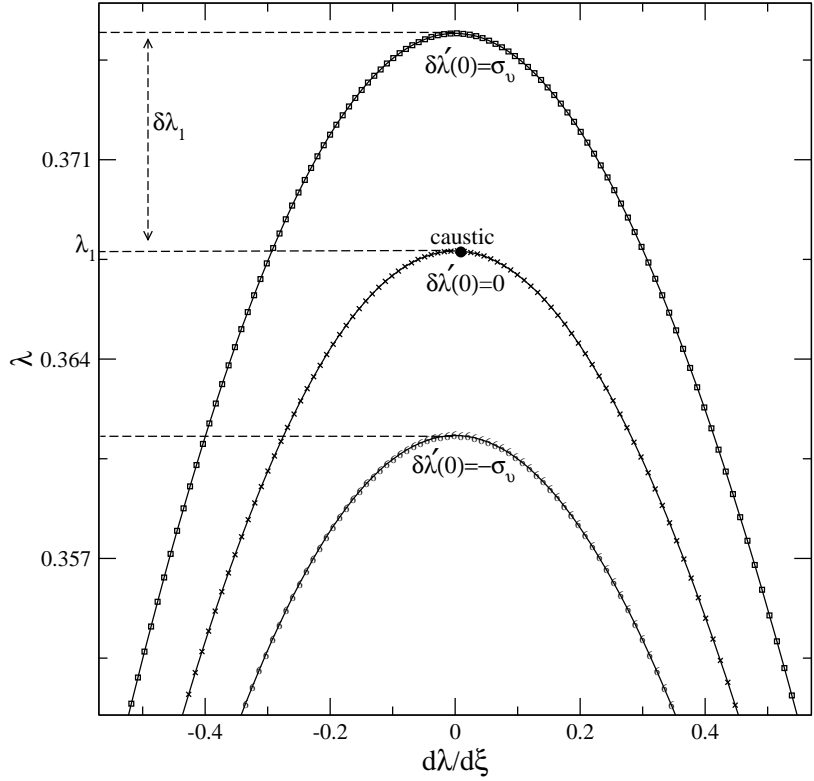}
\ec
\caption
{
The flipped plot of the phase space around the first caustic 
for three values of the initial velocity
perturbation, $\sim\delta\lambda_0$. Phase plots are used to evaluate
the density at the caustics in the presence of a small velocity dispersion.
}
\label{vlambdapeaks-20march2004-flipped}
\end{figure}

\begin{figure}
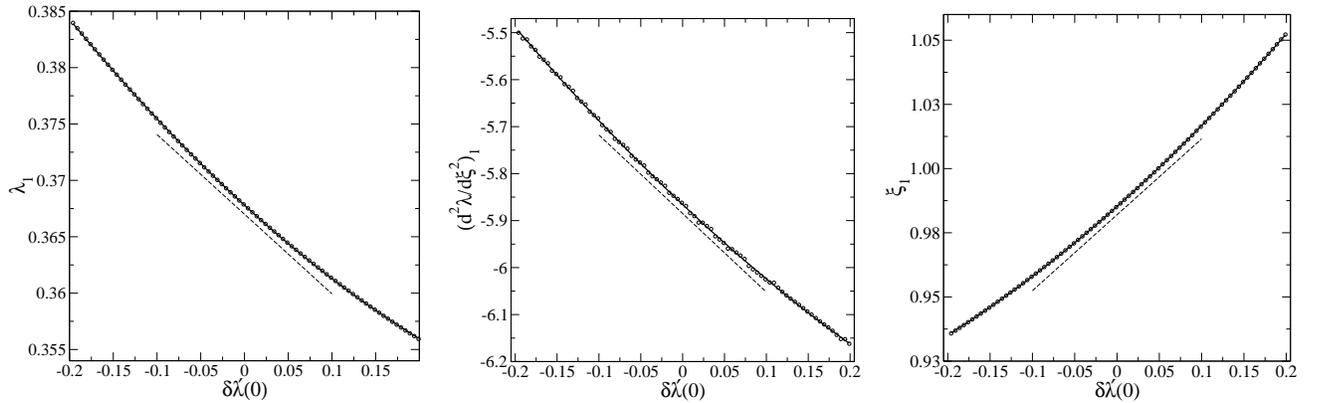

\ba
\includegraphics[width=55mm]{Fig8.eps}
&
\includegraphics[width=55mm]{Fig9.eps}\quad
\includegraphics[width=55mm]{Fig10.eps}
\nonumber\\
\nn
\ea
\caption
{
The left plot shows the variation of the 
non-dimensional position of the first caustic with 
the initial velocity perturbation, $\delta\lambda^\prime(0)$. The
middle and right panels show the variations of the second derivative and the
time of formation of the first caustic with velocity dispersion, respectively. 
The quadratic fits (solid line) are also shown. The dashed straight lines
demonstrate that for very small velocity dispersion, a linear function would
be equally appropriate. Similar fits were obtained for the first ten caustics.
}
\label{lambda02u}
\end{figure}

We consider the process in the nondimensional coordinates
$\xi, \, \lambda, \, \lambda^{\prime} d\lambda/d\xi$.
As we have mentioned in Subsection \ref{sec:Med_with_Therm_Vel}, 
the effects of thermal velocity dispersion can be considered by
adding small velocities 
$ \delta\lambda^{\prime}_0=\delta\lambda^{\prime}(0)$
to the initial velocity. Thus, the initial conditions become
\be
\lambda_0 \equiv \lambda(0)=1, \quad
\lambda^{\prime}_0 \equiv \lambda^{\prime}(0)
=- \frac{8}{9} + \delta \lambda^{\prime}_0.
\label{eq:lambda_dot}
\ee
The result of the integration is shown 
in Fig. \ref{vlambdapeaks-20march2004-flipped} for the first
caustic.
For small values of $\delta \lambda^{\prime}_0$, 
the major effect on the caustic is the change of the maximum value 
$\lambda_k$ that can be well-approximated by a linear function
\be
\delta \lambda_k= \Lambda_{k} \delta\lambda^{\prime}_0
\label{eq:coeff}
\ee
as is evident from Fig. \ref{lambda02u}.
We find $\Lambda_k$ by fitting the numerical results
for every caustic $k=1,...,10$.

%-------------------------------------------------------------------------
%-----------------------------------------------------------------------------

%---------------------------------------------------------------------------
%------------------------------------------------------------------------ 

%%%%%%%%%%%%%%%%%%%%%%%%%%%%%%%%%%%%%%%%%%%%%%%%%%%%%%%%%%%%%%%%%
\section{The density profiles for the top-hat, exponential and
Gaussian velocity distributions}
\label{appendix:exp_gauss}
%----------------------------------------------------------------
We evaluated the density profile in the vicinity of a caustic
for the top-hat initial thermal velocity distribution 
\be
f_{TH}(v)=\rho_0\cases{ \frac{1}{2\sigma_v} ~ \qquad
\qquad \quad |v| < \sigma_v \cr  0\qquad\qquad\quad ~{\rm otherwise} \cr}
\ee
in Sec. \ref{sec:Med_with_Therm_Vel} [equation (\ref{eq:den_TH})].
Using similar approach one also can derive the density profiles 
for the exponential 
and Gaussian velocity distributions
\ba
f_{E}(v)&=&\rho_0\frac{1}{2\sigma_v} \exp{\left( -\frac{|v|}{\sigma_v}\right )}
\\
f_{G}(v)&=&\rho_0\frac{1}{\sqrt{2\pi} \sigma_v} 
\exp{\left (-\frac{v^2}{2\sigma_v^2}\right )}.
\ea

Thus, deriving the density profile in the vicinity of a caustic in both cases
consists in straightforward  evaluations of a few integrals. However,
one can simplify the calculations by
introducing the scaled distance from the caustic $\epsilon$ 
and the scaled density $\eta$
\be
\epsilon=\frac{\Delta x}{|\alpha_k\sigma_v|}, \quad
\eta=\frac{\rho}{A_k\left|\alpha_k\sigma_v\right|^{-1/2}}
\label{eq:eps_eta}
\ee

In terms of these variables  
the density of cold dark matter with zero velocity dispersion 
has a simple form
\be
\eta_{\sigma_v=0}=\cases{
(-\epsilon)^{-1/2} \ \ \qquad \mathrm{for}\ \epsilon<0 \cr\cr
\ \mathrm{\ 0\ \qquad \qquad\quad \mathrm{for} \ \epsilon>0}\cr},
\label{etacold}
\ee
that can be obtained as a limit at $\sigma_v \rightarrow 0$ of any  of three
expressions bellow.

For the top-hat velocity distribution function, with non-vanishing
$\sigma_v$, we have
\be
\eta_{{\rm TH}}=\cases{ 
\sqrt{1 -\epsilon} - \sqrt{-1 -\epsilon}\ \qquad\quad\quad \ \ \mathrm{for}\ \quad 
\epsilon \le -1,
\cr\cr
\sqrt{1-\epsilon} \ \ \qquad\qquad\qquad\qquad\quad \mathrm{for}\ -1 \le \epsilon \le 1,
\cr\cr
0  \qquad\qquad\qquad\qquad\qquad\qquad\quad \mathrm{for} \ \ \epsilon \ge 1.
\cr
}
\label{etaTH}
\ee
For the exponential velocity distribution function one obtains
\be
\eta_{{\rm E}} = 
\cases{
\frac{\sqrt{\pi}}{2}\left\{ \mathrm{e}^{-\epsilon}\,
\left[1-\mathrm{erf}(\sqrt{-\epsilon})\right]
-\mathrm{i}\,\mathrm{e}^{\epsilon} \,\mathrm{erf}(\mathrm{i}\sqrt{-\epsilon})\right\} \ \ \qquad\qquad
\mathrm{for}\quad\ \epsilon<0 
\cr\cr
\frac{\sqrt{\pi}}{2} \mathrm{e}^{-\epsilon} \qquad \qquad \qquad\qquad \qquad
\qquad \qquad \qquad \qquad \qquad \ \ \mathrm{for}\ \quad \epsilon>0
\cr}
\label{etaE}
\ee
and finally for the Gaussian velocity distribution function the
density is as follows
\be
\eta_{{\rm G}}=\cases{
\sqrt{\left|\frac{\epsilon}{8\pi}\right|}\mathrm{e}^{-\epsilon^2/4}\left[
2\pi\mathrm{BesselI}\left(-\frac{1}{4},\frac{\epsilon^2}{4}\right)
-\sqrt{2}\mathrm{BesselK}\left(\frac{1}{4},\frac{\epsilon^2}{4}\right)\right] \ 
\qquad \qquad \mathrm{for}\ \epsilon \le 0 \cr\cr
\sqrt{\left|\frac{\epsilon}{4\pi}\right|}\mathrm{e}^{-\epsilon^2/4}
\mathrm{BesselK}\left(\frac{1}{4},\frac{\epsilon^2}{4}\right)\ \qquad \qquad\qquad
\qquad\qquad \qquad \qquad \qquad\mathrm{for}\ \epsilon \ge 0. \cr}
\label{etaG}
\ee
The density profiles $\eta=\eta(\epsilon)$ are
shown in Fig. \ref{fig:den_prof}.

We also show a simple approximation used in
further calculations
\be
\eta_{{\rm A}}=\cases{ 
(-\epsilon)^{-1/2} \qquad\quad\quad \ \ \mathrm{for}\ \quad 
\epsilon \le -1,
\cr\cr
1 \ \ \qquad\qquad\qquad\quad \mathrm{for}\ -1 \le \epsilon \le 0,
\cr\cr
0 \ \ \qquad\qquad\qquad\quad \mathrm{for} \ \ \epsilon > 0.
\cr}
\label{eta_model}
\ee


\begin{thebibliography}{}

\frenchspacing

%%%%%%%%%%%%%%%%%%%%%%%%%%%%%%%%%%%%%%%%%%
\bibitem{ca2004}
Alard C., Colombi S., {\it A cloudy Vlasov solution}, astro-ph/0406617
%%%%%%%%%%%%%%%%%%%%%%%%%%%%%%%%%%%%%%%%%%
\bibitem{asz82}
Arnol'd V.I., Shandarin S., Zel'dovich Ya.-B, 1982,
Geophysical and Astrophysical Fluid Dynamics, {\bf 20}, 111
%%%%%%%%%%%%%%%%%%%%%%%%%%%%%%%%%%%%%%%%%%
\bibitem{Arnol'd86}
Arnol'd V.I., 1986, {\it Catastrophe Theory}, 
Springer-Verlag Telos, 2nd Edition
%%%%%%%%%%%%%%%%%%%%%%%%%%%%%%%%%%%%%%%%%%
\bibitem{arn-gz-var85}
Arnol'd V.I., Gusein-Zade S.M., Varchenko A.N. 1985, Singularities of Differentiable
Maps, Birkhauser Boston, Cambridge, Massachussets
%%%%%%%%%%%%%%%%%%%%%%%%%%%%%%%%%%%%%%%%%%
\bibitem{Arnol'd1990}
Arnol'd V.I., 1990, {\it Singularities of caustics and wave fronts},
Kluwer Academic publishers, Mathematics and its applications (Soviet Series)
volume 62.
%%%%%%%%%%%%%%%%%%%%%%%%%%%%%%%%%%%%%%%%%%
\bibitem{beg2001}
Bergstr$\ddot {\rm o}$m L., Edsj$\ddot {\rm o}$, Gunnarsson C. 
2001, Phys. Rev D {\bf 63}, 083515
%%%%%%%%%%%%%%%%%%%%%%%%%%%%%%%%%%%%%%%%%%
\bibitem[\protect\citename{Bertschinger} {1985a}]{bert85a}
Bertschinger E. 1985a, ApJ {\bf 58}, 1
%%%%%%%%%%%%%%%%%%%%%%%%%%%%%%%%%%%%%%%%%%
\bibitem[\protect\citename{Bertschinger} {1985b}]{Bert85b}
Bertschinger E. 1985b, ApJ {\bf 58}, 39
%%%%%%%%%%%%%%%%%%%%%%%%%%%%%%%%%%%%%%%%%%
\bibitem{bsssy2000}
Bharadwaj S.,Sahni V., Sathyaprakash B.S., Shandarin S.F., Yess C., 2000,
ApJ {\bf 528}, 21  
%%%%%%%%%%%%%%%%%%%%%%%%%%%%%%%%%%%%%%%%%%
\bibitem{binney2004}
Binney J., 2004, MNRAS {\bf 350}, 939
%%%%%%%%%%%%%%%%%%%%%%%%%%%%%%%%%%%%%%%%%%
\bibitem{hogan2001}
Hogan C., 2001, Phys. Rev. D {\bf 64}, 063515
%%%%%%%%%%%%%%%%%%%%%%%%%%%%%%%%%%%%%%%%%%
\bibitem[die_moo_sta05]{}
Diemand J., Moore B., Stadel J., 2005, Nature {\bf 433}, 389
%%%%%%%%%%%%%%%%%%%%%%%%%%%%%%%%%%%%%%%%%%
\bibitem[91]{}
Dubinski J., Carlberg R., 1991, ApJ {\bf 378}, 496
%%%%%%%%%%%%%%%%%%%%%%%%%%%%%%%%%
\bibitem{efs2003}
Evans N.W., Ferrer F., Sarkar S., 2004, Phys. Rev. D {\bf 69}, 123501
%%%%%%%%%%%%%%%%%%%%%%%%%%%%%%%%%%%
\bibitem{falvardetal2002}
Falvard A. Giraud E., Jacholkowska A., Lavalle J., Nuss E., Piron F., Sapinski M.,
Salati P., Taillet R., Jedamzik K., Moultaka G. 2004, Atropart. Phys. {\bf 20}, 467
%%%%%%%%%%%%%%%%%%%%%%%%%%%%%%%%%%%%%%%%%%%
\bibitem[\protect\citename{Filmore \& Goldreich} 1984]{fg84}
Fillmore J.A., Goldreich P. 1984, ApJ {\bf 281}, 1
%%%%%%%%%%%%%%%%%%%%%%%%%%%%%%%%%%%%%%%%%%%
\bibitem{gondoloandsilk}
Gondolo P., Silk J., 1999, Phys. Rev. Lett. {\bf 83}, 1719 
%%%%%%%%%%%%%%%%%%%%%%%%%%%%%%%%%%%%%%%%%%%
\bibitem[75]{}
Gott J.R., 1975, ApJ {\bf 201}, 296
%%%%%%%%%%%%%%%%%%%%%%%%%%%%%%%%%%%%%%%%%%%
%%%%%%%%%%%%%%%%%%%%%%%%%%%%%%%%%%%%%%%%%%%
\bibitem{gunn77}
Gunn J.E. 1977, ApJ {\bf 218}, 592
%%%%%%%%%%%%%%%%%%%%%%%%%%%%%%%%%%%%%%%%%%%
\bibitem{helmi2003}
Helmi A., White S.D.M., Springel V. 2003, MNRAS {\bf 339} 834
%%%%%%%%%%%%%%%%%%%%%%%%%%%%%%%%%%%%%%%%%%%
\bibitem{hq88a}
Henriksen R.N. 2004, MNRAS {\bf 355}, 1217
%%%%%%%%%%%%%%%%%%%%%%%%%%%%%%%%%%%%%%%%%%%
\bibitem{hog01}
Hogan C. 2001, Phys.Rev. D {\bf 64}, 063515
%%%%%%%%%%%%%%%%%%%%%%%%%%%%%%%%%%%%%%%%%%%
\bibitem{jkg1996}
Jungman G., Kamionkowski M., Griest K., 1996, Phys. Rep. {\bf 267}, 195
%%%%%%%%%%%%%%%%%%%%%%%%%%%%%%%%%
\bibitem{karachentsev2002}
Karachentsev I.D. et al., 2002, A\& A {\bf 389}, 812
%%%%%%%%%%%%%%%%%%%%%%%%%%%%%%%%%%%%%%%%%%%
\bibitem{klypin}
Klypin A., Kravtsov A.V., Valenzuela O., Prada F., 1999, ApJ {\bf 52}, 82
%%%%%%%%%%%%%%%%%%%%%%%%%%%%%%%%%%%%%%%%%%%
\bibitem{kotok}
Kotok E.V., Shandarin S.F. 1987, Sov. Astron. 31, 600
%%%%%%%%%%%%%%%%%%%%%%%%%%%%%%%%%%%%%%%%%%%
\bibitem{mcgaugh2003}
McGaugh S.S, Barker M.K., de Blok W.J.G., 2003, ApJ {\bf 584}, 566
%%%%%%%%%%%%%%%%%%%%%%%%%%%%%%%%%%%%%%%%%%%
\bibitem{melott1998}
Melott A.L., Shandarin S.F., Splinter R.J., Suto Y., 1997, ApJL {\bf 479}, 79 
%%%%%%%%%%%%%%%%%%%%%%%%%%%%%%%%%%%%%%%%%%%
\bibitem{moore1}
Moore, B., Ghigna S., Governato F., Lake G., Quinn T., Stadel J., Tozzi P.,
1999, ApJL {\bf 524}, 19
%%%%%%%%%%%%%%%%%%%%%%%%%%%%%%%%%%%%%%%%%%%
\bibitem{moore1}
B. Moore, "Caustic rings and cold dark matter", in the Proceedings
  of the 3d International Workshop on the Identification of Dark Matter,
  York, UK, September 18-22, 2000, edited by N.J.C. Spooner and
  V. Kudryatsev, World Scientific 2001, p93
\bibitem[96]{}
Navarro J.F., Frenk C.S., White S.D.M., 1996, ApJ {\bf 62}, 563
%%%%%%%%%%%%%%%%%%%%%%%%%%%%%%%%%%%%%%%%%%%
\bibitem[97]{}
Navarro J.F., Frenk C.S., White S.D.M., 1997, ApJ {\bf 490}, 493
%%%%%%%%%%%%%%%%%%%%%%%%%%%%%%%%%%%%%%%%%%%
\bibitem{nussetal2002}
Nuss E. et al. ,in Semaine de l'Astrophysique Francaise, Paris,
France, June 24-29, 2002, Eds.: F. Combes and D. Barret, EdP-Sciences
(Editions de Physique), 279
%%%%%%%%%%%%%%%%%%%%%%%%%%%%%%%%%%%%%%%%%%%
\bibitem{roytvarf}
Roytvarf A., 1994, Physica D {\bf 73}, 189
%%%%%%%%%%%%%%%%%%%%%%%%%%%%%%%%%%%%%%%%%%%
\bibitem{sahni1996}
Sahni V., Shandarin, S., 1996, MNRAS {\bf 282}, 641 
%%%%%%%%%%%%%%%%%%%%%%%%%%%%%%%%%%%%%%%%%%%
\bibitem{salati2004}
Salati P., 2004, in 4\`eme semaine de l'astrophysique francaise, Paris,
eds: F. Combes et al., EDP Sciences
%%%%%%%%%%%%%%%%%%%%%%%%%%%%%%%%%%%%%%%%%%%
\bibitem{sandage1986}
Sandage A., 1986, ApJ {\bf 307}, 1
%%%%%%%%%%%%%%%%%%%%%%%%%%%%%%%%%%%%%%%%%%%
\bibitem{sz89}
Shandarin S.F., Zel'dovich Ya.B. 1989, 
Rev. Mod. Phys. {\bf 61}, 185
%%%%%%%%%%%%%%%%%%%%%%%%%%%%%%%%%%%%%%%%%%%
\bibitem{stw97}
Sikivie P. 1999, Phys. Rev. D {\bf 60}, 063501
%%%%%%%%%%%%%%%%%%%%%%%%%%%%%%%%%%%%%%%%%%%
\bibitem{stw97}
Sikivie P., Ipser J. 1992, Phys. Lett. B {\bf 291}, 288
%%%%%%%%%%%%%%%%%%%%%%%%%%%%%%%%%%%%%%%%%%%
\bibitem{stw97}
Sikivie P., Tkachev I.I., Wang Y. 1997, Phys. Rev. D {\bf 56}, 1863
%%%%%%%%%%%%%%%%%%%%%%%%%%%%%%%%%%%%%%%%%%%
\bibitem{splinter1998}
Splinter R.J., Melott A.L., Shandarin S.F., Suto Y., 1998, ApJ {\bf 497}, 38
%%%%%%%%%%%%%%%%%%%%%%%%%%%%%%%%%%%%%%%%%%%
\bibitem{swsty2003}
Stoehr F., White S.D.M., Springel V., Tormen G., Yoshida N., 2003, MNRAS {\bf 345}, 1313
%%%%%%%%%%%%%%%%%%%%%%%%%%%%%%%%%%%%%%%%%%%
\bibitem{stw97}
Tremaine S. 1999, MNRAS {\bf 307}, 877
%%%%%%%%%%%%%%%%%%%%%%%%%%%%%%%%%%%%%%%%%%%
\bibitem[70]{}
Zel'dovich Ya.B., 1970, A\&A {\bf 5}, 84
%%%%%%%%%%%%%%%%%%%%%%%%%%%%%%%%%%%%%%%%%%%
\bibitem[70]{}
Zel'dovich Ya.B., Shandarin S.F. 1982, Soviet Ast. Lett. {\bf 8}, 139
%%%%%%%%%%%%%%%%%%%%%%%%%%%%%%%%%%%%%%%%%%%


\end{thebibliography}
\end{document}